\documentclass[journal,comsoc]{IEEEtran}
\usepackage[T1]{fontenc}
\usepackage{longtable}
\usepackage{supertabular} 
\usepackage{graphicx}
\usepackage{lineno,hyperref}
\usepackage{amsmath}
\usepackage{amsfonts}
\usepackage{amssymb}
\usepackage{array,multirow,makecell}
\modulolinenumbers[5]
\usepackage{lscape}
\usepackage[cmintegrals]{newtxmath}
\usepackage[table]{xcolor}
	
\begin{document}
%
%
%
%
\title{Security for 4G and 5G Cellular Networks: A Survey of Existing Authentication and Privacy-preserving Schemes}
\author{Mohamed~Amine~Ferrag,
        Leandros~Maglaras,
        Antonios~Argyriou,
        Dimitrios~Kosmanos,
        and Helge~Janicke
\thanks{(Corresponding author: Mohamed Amine Ferrag)}
\thanks{M. A. Ferrag is with Department of Computer Science, Guelma University, 24000, Algeria, and also with Networks and Systems Laboratory (LRS), Badji Mokhtar-Annaba University, 23000 Annaba, Algeria e-mail: mohamed.amine.ferrag@gmail.com  phone: +213661-873-051}
\thanks{L. Maglaras and H. Janicke are with School of Computer Science and Informatics, De Montfort University, Leicester, UK e-mails: \{leandros.maglaras, heljanic\}@dmu.ac.uk}
\thanks{A. Argyriou and D. Kosmanos are with Department of Electrical and Computer Engineering, University of Thessaly, Greece e-mails: anargyr@gmail.com, dimitriskosmanos@gmail.com}
\thanks{Manuscript received 2017.}}

\maketitle
\begin{abstract}
This paper presents a comprehensive survey of existing authentication and privacy-preserving schemes for 4G and 5G cellular networks. We start by providing an overview of existing surveys that deal with 4G and 5G communications, applications, standardization, and security. Then, we give a classification of threat models in 4G and 5G cellular networks in four categories, including, attacks against privacy, attacks against integrity, attacks against availability, and attacks against authentication. We also provide a classification of countermeasures into three types of categories, including, cryptography methods, humans factors, and intrusion detection methods. The countermeasures and informal and formal security analysis techniques used by the authentication and privacy preserving schemes are summarized in form of tables. Based on the categorization of the authentication and privacy models, we classify these schemes in seven types, including, handover authentication with privacy, mutual authentication with privacy, RFID authentication with privacy, deniable authentication with privacy, authentication with mutual anonymity, authentication and key agreement with privacy, and three-factor authentication with privacy. In addition, we provide a taxonomy and comparison of authentication and privacy-preserving schemes for 4G and 5G cellular networks in form of tables. Based on the current survey, several recommendations for further research are discussed at the end of this paper.
\end{abstract}

\IEEEpeerreviewmaketitle
\section{Introduction}

\begin{figure}
\centering
\includegraphics[width=1\linewidth]{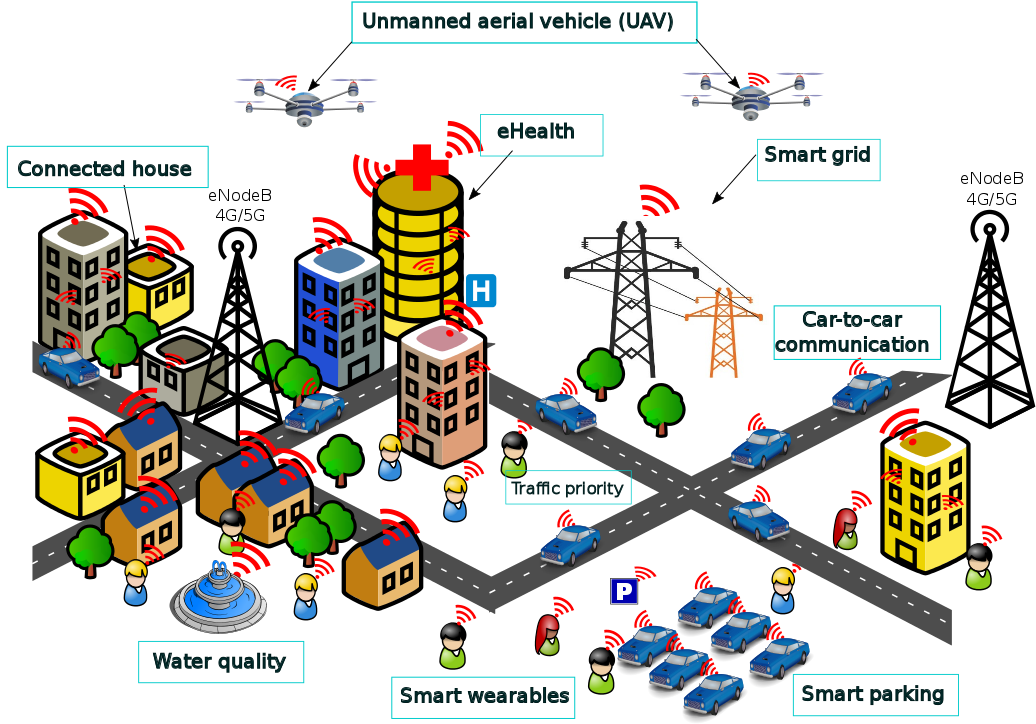}
\caption{What will 5G enable?}
\label{fig:Fig1_1}
\end{figure}

The fifth-generation mobile networks (5G) will soon supersede 4G in most countries of the world. The next generation wireless network technology is being developed based on recent advances in wireless and networking technologies such as software-defined networking and virtualization. Compared to 4G technologies, 5G is characterized by still higher bit rates with more than 10 gigabits per second as well as by more capacity and very low latency, which is a major asset for the billions of connected objects in the context of Internet of Things (IoT). In the IoT era, 5G will enable a fully mobile and connected society, via creating various new network services such as mobile fog computing, car-to-car communications, smart grid, smart parking, named data networking, blockchain based services, unmanned aerial vehicle (UAV) etc. as shown in Fig. \ref{fig:Fig1_1}. Therefore, telecommunications companies believe that the commercialization of 5G will begin in 2020. In Tab. \ref{Table:Tab1_1}, we list some of the leading projects for 5G cellular networks by various telecommunications companies.
\begin{table}[!t]
	\centering
	\caption{The leading projects for 5G}
	\label{Table:Tab1_1}
	\setlength{\tabcolsep}{2pt}
		\vspace*{-\baselineskip}
		\renewcommand{\arraystretch}{1.5}
{\tiny
\begin{tabular}{p{0.2in}|p{1in}|p{2in}} \hline 
\textbf{Time} & \textbf{Company} & \textbf{Program} \\ \hline \hline
2014 & NTT DOCOMO and SK Telecom & Ericsson 5G delivers 5 Gbps speeds \cite{L1} \\ \hline 
2016 & Ericsson and SoftBank & Ericsson and SoftBank completed basic 5G trials on both 15 GHz and 4.5GHz spectrums \cite{L7} \\ \hline 
2016 & Ericsson and Telef\'{o}nica & Ericsson and Telef\'{o}nica focused on the Advanced 5G Network Infrastructure for Future Internet Public-Private Partnership (5G PPP) and European Technology Platform for Communications Networks and Services (ETP Networld 2020) \\ \hline 
2016 & Huawei  and Vodafone Group plc & Vodafone Group with Huawei have recently completed a 5G field test in Newbury (UK) that demonstrates the capabilities of a trial system operating at 70 GHz \cite{L5} \\ \hline 
2017 & Huawei and China Mobile Ltd. & Huawei and China Mobile showcased the 5G 3.5GHz prototype and Ka-Band millimeter wave prototype \cite{L2} \\ \hline 
2017 & Verizon Communications Inc. & Verizon will begin pilot testing 5G "pre-commercial services" in U.S. cities in the first half of 2017, including Atlanta, Dallas, Denver, Houston, Miami, Seattle, and Washington \cite{L3} \\ \hline 
2017 & AT\&T Inc. & AT\&T launches Nationwide LTE-M Network for Internet of Things \cite{L4} \\ \hline 
2017 & Nippon Telegraph and Telephone Corporation (NTT) & Toyota and NTT collaborate to promote 5G standardization for automotive vehicles \cite{L6} \\ \hline 
2017 & Huawei and Deutsche Telekom & Huawei and Deutsche Telekom demonstrate the all Cloud 5G network slicing \cite{L2} \\ \hline 
\end{tabular}}
\end{table}
\begin{table*}[!t]
	\centering
	\caption{Definitions of acronyms and notations}
	\label{Table:Tab1_2}
	\setlength{\tabcolsep}{2pt}
		\vspace*{-\baselineskip}
		\renewcommand{\arraystretch}{1.5}
{\scriptsize
\begin{tabular}{p{0.6in}p{2.3in}|p{0.6in}p{2in}} \hline 
\textbf{Acronym } & \textbf{Definition} & \textbf{Acronym} & \textbf{Definition} \\ \hline \hline
\textbf{}3GPP & Third Generation Partnership Project & IRS & Intrusion Response System \\ 
\textbf{}4G & Fourth-generation mobile network & LTE & Long-Term Evolution \\ 
\textbf{}5G & Fifth-generation mobile network & LTE-A & Long-Term Evolution Advenced \\   
\textbf{}AES & Advanced Encryption Standard & M2M & Machine-to-Machine \\   
\textbf{}AIM & Advanced Identity Management & MAC & Message Authentication Code \\   
\textbf{}AKA & Authentication and Key Agreement & MD5 & Message Digest 5 \\   
\textbf{}AMAC & Aggregate Message Authentication Codes & MIMO & Multiple-Input Multiple-Output \\   
\textbf{}AP & Access point & MITM & Man-in-the-middle \\   
\textbf{}BRPCA & Bayesian Robust Principal Component Analysis & MME & Mobility Management Entity \\   
\textbf{}BS & Base station & MSS & Managed security services \\   
\textbf{}BTS & Base Transceiver Station & MTC & Machine Type Communication \\   
\textbf{}CNN & Controller Area Network & NB & Narrowband \\   
\textbf{}CRC & Cyclic Redundancy Check & NFV & Network Function Virtualization \\   
\textbf{}CXTP & Context transfer protocol & P2P & Peer-to-Peer \\   
\textbf{}D2D & Device-to-Device communication & PIN & Personal identification number \\   
\textbf{}DNN & Deep Neural Network & PKI & Public key infrastructure \\   
\textbf{}DoS & Denial of Service & PT  & Pseudo Trust \\   
\textbf{}DSS & Digital signature standard & RAN & Radio Access Network \\   
\textbf{}EAP & Extensible Authentication Protocol & RF & Radio Frequency \\   
\textbf{}ECC & Error Correction Codes & RFC & Requests For Comments \\   
\textbf{}eNB & eNodeB & RFID & Radio frequency identification \\   
\textbf{}FBS & False Base Station & RNN & Random Neural Network \\   
\textbf{}FIFO & First In First Out & RNTI & Radio Network Temporary Identities \\   
\textbf{}GBS-AKA & Group-Based Secure Authentication and Key Agreement & SDN & Software Defined Networking \\   
\textbf{}HeNB & Home eNodeB & SHA & Secure Hash Algorithm \\   
\textbf{}HMAC & Keyed-Hash Message Authentication Code & SIP & Session Initiation Protocol \\   
\textbf{}HSS & Home Service Server & TLS & Transport Layer Security \\   
\textbf{}HTTP & Hypertext Transfer Protocol & TPM & Trusted Platform Module \\   
\textbf{}IDS & Intrusion Detection system & UAV & Unmanned aerial vehicle \\   
\textbf{}IEEE & Institute of Electrical and Electronics Engineers & UE & User Equipment \\   
\textbf{}IMSI & International Mobile Subscriber Identity & UHF & UltraHigh Frequency \\   
\textbf{}IoT & Internet of Things & UMTS & Universal Mobile Telecommunications System \\ \hline 
\end{tabular}}
\end{table*}

In a 5G environment, the blend different wireless technologies and service providers that share an IP-based core network, will offer the possibility to the mobile devices of switching between providers and technologies, for maintaining a high level of Quality of Service (QoS). Fast vertical handover and the general openness of the network make the devises susceptible to several vulnerabilities like access control, communication security, data confidentiality, availability and privacy. Furthermore, since the 5G environment is IP-based, it will suffer from all the vulnerabilities that are to IP-specific.  Based on these findings, it is obvious that guaranteeing a high level of security and privacy will be one of important aspects for the successful deployment of 5G networks \cite{31}.

As mobile devices will be connected to the network all the time, through the vertical handover, they will obtain a notion of social nodes. Such nodes can more easily be tracked down and are more vulnerable in several types of attacks, like impersonation, eavesdropping, man-in-the-middle, denial-of-service, replay and repudiation attack \cite{155}. Maintaining a high level of  QoS in terms of delay, when huge volume of data is transferred inside a 5G network, while keeping on the same time  high security and privacy level, is critical in order to prevent malicious files from penetrating the system and propagating fast among mobile devices. Thus  communications that satisfy zero latency requirements are cumbersome once combined with secure and privacy-preserving 5G networks \cite{L21}.

For the process of conducting the literature review, we follow the same process conducted by our previous work in \cite{155}. Specifically, the identification of literature for analysis in this paper was based on a keyword search, namely, "authentication and privacy-preserving scheme", "authentication and privacy-preserving protocol", "authentication and privacy-preserving system", and "authentication and privacy-preserving framework". Searching for these keywords in academic databases such as SCOPUS, Web of Science, IEEE Xplore Digital Library, and ACM Digital Library, an initial set of relevant sources were located. Firstly, only proposed authentication and privacy-preserving schemes for 4G and 5G cellular networks were collected. Secondly, each collected source was evaluated against the following criteria: 1) reputation, 2) relevance, 3) originality, 4) date of publication (between 2005 and 2017), and 5) most influential papers in the field. The final pool of papers consists of the most important papers in the field of 4G and 5G cellular networks that focus on the authentication and privacy-preserving as their objective. Our search started on 15/01/2017 and continued until the submission date of this paper.

The main contributions of this paper are:

\begin{itemize}
\item  We discuss the existing surveys for 4G and 5G cellular networks that deal with communications, applications, standardization, and security.

\item  We provide a classification for the attacks in cellular networks in four categories, including, attacks against privacy, attacks against integrity, attacks against availability, and attacks against authentication.

\item  We provide a classification for countermeasures used by the authentication and privacy preserving schemes for 4G and 5G cellular networks into three types of categories, including, cryptography methods, humans factors, and intrusion detection methods.

\item  We present the informal and formal security analysis techniques used by the authentication and privacy preserving schemes for 4G and 5G cellular networks.

\item  We provide a categorization of authentication and privacy models for 4G and 5G cellular networks.

\item  We provide a classification of authentication and privacy preserving schemes for 4G and 5G cellular networks in seven types, including, handover authentication with privacy, mutual authentication with privacy, RFID authentication with privacy, deniable authentication with privacy, authentication with mutual anonymity, authentication and key agreement with privacy, and three-factor authentication with privacy.

\item  We outline six recommendations for further research, including, 1) privacy preservation for Fog paradigm-based 5G radio access network, 2) authentication for 5G small cell-based smart grids, 3) privacy preservation for SDN/NFV-based architecture in 5G scenarios, 4) dataset for intrusion detection in 5G scenarios, 5) privacy preserving schemes for UAV systems in 5G heterogeneous communication environment, and 6) authentication for 5G small cell-based vehicular crowdsensing.
\end{itemize}

The remainder of this paper is organized as follows. Section \ref{sec:existing-surveys-for-4g-and-5g-cellular-networks} presents the existing surveys for 4G and 5G cellular networks that deal with communications, applications, standardization, and security. In Section \ref{sec:threat-models-and-countermeasures}, we provide a classification for the threat models and countermeasures. Section \ref{sec:informal-and-formal-security-analysis-techniques} presents various the informal and formal security analysis techniques used by the authentication and privacy preserving schemes for 4G and 5G cellular networks. In Section \ref{sec:authentication-and-privacy-preserving-schemes-for-4g-and-5g-cellular-networks}, we present a side-by-side comparison in a tabular form for the current state-of-the-art of authentication and privacy preserving schemes for 4G and 5G cellular networks. Then, we discuss open issues and recommendations for further research in Section \ref{sec:future-directions}. Finally, we draw our conclusions in Section \ref{sec:conclusions}. Table \ref{Table:Tab1_2} lists the acronyms and notations used in the paper.

\section{Existing surveys for 4G and 5G Cellular Networks}\label{sec:existing-surveys-for-4g-and-5g-cellular-networks}
There are around fifty survey articles published in the recent years that deal with 4G and 5G communications, applications, standardization and security. These survey articles are categorized as shown in tables \ref{Table:Tab1} and \ref{Table:Tab2}. From these survey articles only seven of them deal with security and privacy issues for 3G, 4G and 5G cellular networks and none of the previous works covers the authentication and privacy preserving issues of 4G and 5G networks. This work is the first on the literature that thoroughly covers authentication and privacy preservation threat models, countermeasures and schemes that we recently proposed from the research community.

For these fifty survey articles that were retrieved from SCOPUS and Web of Science and were published from 2007 to 2017 we performed a categorization which is presented in table \ref{Table:Tab1}. Based on this categorization it is obvious that except from three big categories of articles, one dealing with scheduling and interference mitigation \cite{1,4,14,19,22,31}, the other with D2D Communication \cite{13,16,26,34,44,47} and the third with security and privacy issues \cite{2,9,11,21,31,49,50}, all areas of research that are somehow related to 3G, 4G and 5G networks were surveyed and presented in previous surveys from at least one review article. As the technology progress and the networks evolve from 3G to 4G, 5G and even 6G \cite{298}, the number of articles that survey 4G and 5G networks increases from only one that was published back in 2007, to over twenty articles published in 2016. This increase on the number reveals an increase on the importance that researchers from around the world give on the new technology and the issues that arise regarding standardization \cite{7,39,48}, mobile internet applications \cite{8}, resource and mobility management \cite{12,38}, energy \cite{28}, MIMO techniques \cite{38,44,46}, social perspectives \cite{35} and so on (See Table \ref{Table:Tab1} for detailed categorization).

Among the aforementioned surveys, the security and privacy issues that are related to the 4G and 5G networks were thoroughly covered and analyzed in previous works \cite{2,9,11,21,31,49,50}.~As it is shown in Tab. \ref{Table:Tab3} authentication and privacy preservation was only covered partially from Cao et al. \cite{11} while the rest of the articles did not cover this major security aspect.~In this article we survey authentication and privacy preserving protocols for 4G/5G networks. Based on this thorough analysis open issues and future directions are identified, that combine both innovative research and novel implementations, along with application of properly adapted existing solutions from other fields. We believe that this study will help researchers focus on the important aspects of authentication and privacy preservation issues in the 4G and 5G area and will guide them towards their future research.

\begin{table*}[!t]
	\centering
	\caption{Areas of research of each survey article for 4G and 5G Cellular Networks}
	\label{Table:Tab1}
	\setlength{\tabcolsep}{2pt}
		\vspace*{-\baselineskip}
		\renewcommand{\arraystretch}{1.5}
		SIM: Scheduling and Interference Mitigation; SP: Security and Privacy; HD: Heterogeneous Deployments; VN: Vehicular Networking; GCN: Green Cellular Networks; STD: Standardization; MIA: Mobile Internet Applications; RC: Random access channel; D2D: Device-to-Device Communication; RMM: Resource \& Mobility Management; DO: Data Offloading; HM: Handover Management; SDN: Software-defined networking; US: Unlicensed Spectrum; ENE: Energy; BN: Backhaul network; DNMA: Downlink Non-orthogonal Multiple Access; MIMO: Multiple-Input Multiple-Output technologies; Soc: Social perspective; CC: Cloud Computing; mmWave: millimeter wave communications; Archi: Architecture.
{\tiny \begin{tabular}{p{1in}|p{0.2in}|p{0.2in}|p{0.2in}|p{0.2in}|p{0.2in}|p{0.2in}|p{0.2in}|p{0.2in}|p{0.2in}|p{0.2in}|p{0.2in}|p{0.2in}|p{0.2in}|p{0.2in}|p{0.2in}|p{0.2in}|p{0.2in}|p{0.2in}|p{0.2in}|p{0.2in}|p{0.3in}|p{0.2in}} \hline
\textbf{Ref.} & \textbf{SIM} & \textbf{SP} & \textbf{HD} & \textbf{VN} & \textbf{GCN} & \textbf{STD} & \textbf{MIA} & \textbf{RC} & \textbf{RMM} & \textbf{D2D} & \textbf{DO} & \textbf{HM} & \textbf{SDN} & \textbf{US} & \textbf{ENE} & \textbf{BN} & \textbf{DNMA} & \textbf{MIMO} & \textbf{Soc} & \textbf{CC} & \textbf{mmWave} & \textbf{Archi} \\ \hline \hline
\cite{1} \cite{4} \cite{14} \cite{19} \cite{22} \cite{31}& $\surd $ &  &  &  &  &  &  &  &  &  &  &  &  &  &  &  &  &  &  &  &  &  \\ \hline 
\cite{2} \cite{9} \cite{11} \cite{21} \cite{31} \cite{49} \cite{50} & & $\surd $ &  &  &  &  &  &  &  &  &  &  &  &  &  &  &  &  &  &  &  &  \\ \hline 
\cite{3} \cite{25} \cite{30} \cite{36} &  & & $\surd $ &  &  &  &  &  &  &  &  &  &  &  &  &  &  &  &  &  &  &  \\ \hline 
\cite{5} \cite{27} &  &  & & $\surd $ &  &  &  &  &  &  &  &  &  &  &  &  &  &  &  &  &  &  \\ \hline 
\cite{6} \cite{15} \cite{41} &  &  &  & & $\surd $ &  &  &  &  &  &  &  &  &  &  &  &  &  &  &  &  &  \\ \hline 
\cite{7} \cite{39} \cite{48} &  &  &  &  & & $\surd $ &  &  &  &  &  &  &  &  &  &  &  &  &  &  &  &  \\ \hline 
\cite{8} &  &  &  &  &  & & $\surd $ &  &  &  &  &  &  &  &  &  &  &  &  &  &  &  \\ \hline 
\cite{10} \cite{31} &  &  &  &  &  &  & & $\surd $ &  &  &  &  &  &  &  &  &  &  &  &  &  &  \\ \hline 
\cite{12} \cite{38} &  &  &  &  &  &  &  & & $\surd $ &  &  &  &  &  &  &  &  &  &  &  &  &  \\ \hline 
\cite{13} \cite{16} \cite{26} \cite{34} \cite{44} \cite{47} &  &  &  &  &  &  &  &  & & $\surd $ &  &  &  &  &  &  &  &  &  &  &  &  \\ \hline 
\cite{17} &  &  &  &  &  &  &  &  &  & & $\surd $ &  &  &  &  &  &  &  &  &  &  &  \\ \hline 
\cite{18} &  &  &  &  &  &  &  &  &  &  & & $\surd $ &  &  &  &  &  &  &  &  &  &  \\ \hline 
\cite{20} \cite{24} &  &  &  &  &  &  &  &  &  &  &  & & $\surd $ &  &  &  &  &  &  &  &  &  \\ \hline 
\cite{23} &  &  &  &  &  &  &  &  &  &  &  &  & & $\surd $ &  &  &  &  &  &  &  &  \\ \hline 
\cite{28} &  &  &  &  &  &  &  &  &  &  &  &  &  & & $\surd $ &  &  &  &  &  &  &  \\ \hline 
\cite{29} \cite{40} &  &  &  &  &  &  &  &  &  &  &  &  &  &  & & $\surd $ &  &  &  &  &  &  \\ \hline 
\cite{32} \cite{37} &  &  &  &  &  &  &  &  &  &  &  &  &  &  &  & & $\surd $ &  &  &  &  &  \\ \hline 
\cite{33} \cite{44} \cite{46} &  &  &  &  &  &  &  &  &  &  &  &  &  &  &  &  & & $\surd $ &  &  &  &  \\ \hline 
\cite{35} &  &  &  &  &  &  &  &  &  &  &  &  &  &  &  &  &  & & $\surd $ &  &  &  \\ \hline 
\cite{42} \cite{45} &  &  &  &  &  &  &  &  &  &  &  &  &  &  &  &  &  &  & & $\surd $ &  &  \\ \hline 
\cite{43} &  &  &  &  &  &  &  &  &  &  &  &  &  &  &  &  &  &  &  & & $\surd $ &  \\ \hline 
\cite{44} &  &  &  &  &  &  &  &  &  &  &  &  &  &  &  &  &  &  &  &  & & $\surd $ \\ \hline 
\end{tabular}}
\end{table*}
\begin{table}[!t]
	\centering
	\caption{Year of publication}
	\label{Table:Tab2}
	\setlength{\tabcolsep}{2pt}
		\vspace*{-\baselineskip}
		\renewcommand{\arraystretch}{1.5}
{\tiny \begin{tabular}{p{3in}|p{0.3in}} \hline 
\textbf{Ref.} & \textbf{Year} \\ \hline \hline
\cite{49} & 2007 \\ \hline 
\cite{1} \cite{2} \cite{50} & 2010 \\ \hline 
\cite{3} & 2011 \\ \hline 
\cite{6} & 2012 \\ \hline 
\cite{4} \cite{5} \cite{7} \cite{8} \cite{9} & 2013 \\ \hline 
\cite{10} \cite{11} \cite{12} \cite{13} \cite{14} \cite{47} \cite{48} & 2014 \\ \hline 
\cite{15} \cite{16} \cite{17} \cite{18} \cite{19} \cite{38} \cite{43} \cite{44} \cite{45} \cite{46} & 2015 \\ \hline 
\cite{20} \cite{21} \cite{22} \cite{23} \cite{24} \cite{25} \cite{26} \cite{27} \cite{28} \cite{29} \cite{30} \cite{31} \cite{32} \cite{33} \cite{34} \cite{38} \cite{39} \cite{40} \cite{41} \cite{42} & 2016 \\ \hline
 \cite{19} \cite{35} \cite{36} \cite{37} & 2017 \\ \hline 
\end{tabular}}
\end{table}
\begin{table}[!t]
	\centering
	\caption{Comparison of related surveys in the literature (Survey on Security and Privacy for 4G and 5G Cellular Networks)}
	\label{Table:Tab3}
	\setlength{\tabcolsep}{2pt}
		\vspace*{-\baselineskip}
		\renewcommand{\arraystretch}{1.5}
$\surd $ :indicates fully supported; X: indicates not supported; 0: indicates partially supported.
{\tiny \begin{tabular}{p{0.6in}|p{0.1in}|p{0.1in}|p{0.1in}|p{0.3in}|p{0.3in}|p{1.3in}} \hline 
\textbf{Ref.} & \textbf{3G} & \textbf{4G} & \textbf{5G} & \textbf{Authen. } & \textbf{Privacy-preserving} & \textbf{Comments} \\ \hline \hline
Park, Y. and Park, T. (2007) \cite{49} & $\surd $ & X & X & X & X & - Presented some security threats on 4G networks. \\ \hline 
Aiash et al. (2010) \cite{50} & $\surd $ & X & X & 0 & X & - Reviewed the X.805 standard for the AKA protocol. \\ \hline 
Seddigh et al. (2010)  \cite{2}  & $\surd $ & X & X & X & X & - Surveyed the security advances for MAC layer in 4G technologies LTE and WiMAX. \\ \hline 
Bikos and Sklavos (2013) \cite{9} & 0 & $\surd $ & X & X & X & - Presented the cryptographic algorithms for LTE.  \\ \hline 
Cao et al. (2014) \cite{11} & X & $\surd $ & X & 0 & X & - Presented the security architectures and mechanisms specified by the 3GPP standard.  \\ \hline 
Lichtman et al. (2016) \cite{21} & X & $\surd $ & X & X & X & - Surveyed the jamming and spoofing mitigation techniques for LTE. \\ \hline 
Panwar et al. (2016) \cite{31} & X & X & $\surd $ & X & X & - Presented the challenges in security and privacy in 5G networks. \\ \hline 
Our Work & 0 & $\surd $ & $\surd $ & $\surd $ & $\surd $ & - Surveyed the authentication and privacy-preserving schemes for 4G and 5G Cellular Networks. \\ \hline 
\end{tabular}}
\end{table}
\begin{figure*}
\centering
\includegraphics[width=1\linewidth]{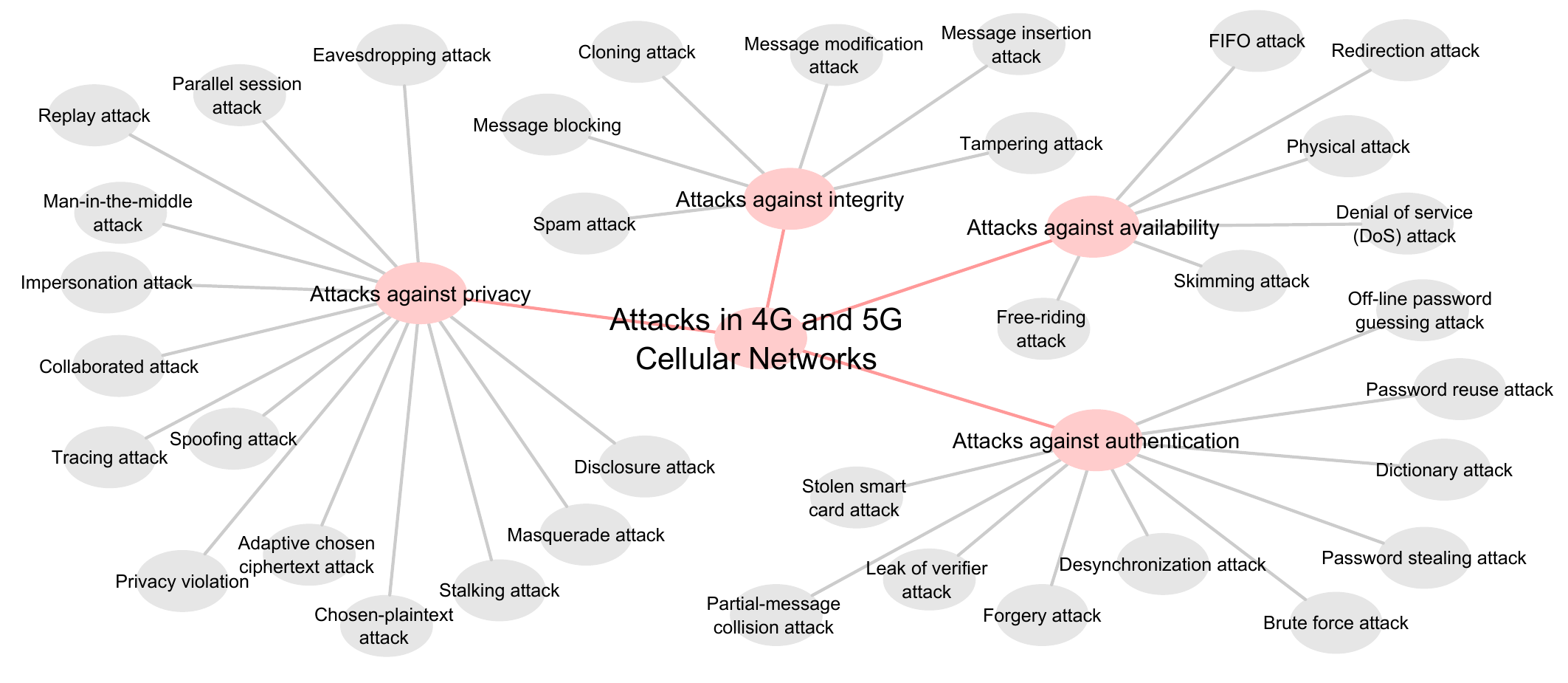}
\caption{Classification of attacks in 4G and 5G Cellular Networks}
\label{fig:Fig3_1}
\end{figure*}
\section{Threat models and countermeasures}\label{sec:threat-models-and-countermeasures}
\subsection{Threat models}\label{sec:threat-models}
In this subsection, we discuss the threat models in 4G and 5G Cellular Networks. We found thirty-five attacks, which are analyzed and prevented by authentication and privacy preserving schemes for 4G and 5G Cellular Networks. The classification of threat models in cellular networks frequently mentioned in literature is done using different criteria such as passive or active, internal or external etc. In our survey article, we classify the attacks in cellular networks in four categories as shown in Fig. \ref{fig:Fig3_1}, including, 1) attacks against privacy, 2) attacks against integrity, 3) attacks against availability, and 4) attacks against authentication. Note that our classification is based on the behavior of the attack in 4G and 5G cellular networks.  

\subsubsection{Attacks against privacy} We classify fourteen attacks in this category, namely, eavesdropping attack, parallel session attack, replay attack, Man-In-The-Middle (MITM) attack, impersonation attack, collaborated attack, tracing attack, spoofing attack, privacy violation, adaptive chosen ciphertext attack, chosen-plaintext attack, stalking attack, masquerade attack, and disclosure attack. The most serious attack among them is the MITM attack. According to Conti et al. \cite{317}, the MITM attack in cellular networks is based on False Base Station (FBS) attack, when malicious third party masquerades its Base Transceiver Station (BTS) as a real network's BTS. Using a temporary confidential channel, Chen et al. \cite{91} proposed an idea that only requires minimum number of human interaction for detecting and avoiding the MITM attack in cellular networks. Mayrhofer et al. \cite{93} proposed a unified cryptographic authentication protocol framework to use with arbitrary auxiliary channels in order to detect the MITM attack in cellular networks. Based on the combination of learning parity with noise, circulant matrix, and multivariate quadratic, Li et al. \cite{94} introduced an entity authentication protocol, which is proved that it is secure against all probabilistic polynomial-time adversaries under MITM attack model. However, note that the MITM attack is a particular case of a replay attack. By support mutual authentication, Chen et al. \cite{97} proposed the improved smart-card-based password authentication and key agreement scheme that can easily detect a replay attack by checking the timestamp. The question we ask here is: Does detecting the replay attack is sufficient to detect the MITM attack? The privacy-preserving authentication scheme proposed recently by Haddad et al. \cite{102} can answer this question where he can prove that the idea of checking the timestamp to detecting the MITM attack is not sufficient, but it is necessary to use the private keys that are not known to the attackers. Yao et al. \cite{122} proposed a group-based secure authentication scheme, named, GBS-AKA, which he can detect the MITM attack using the session keys and timestamp during the authentication procedure. Through the MITM attack, the attacker can launch the other attacks of this category such as eavesdropping attacks to intercept keys and messages by unintended receivers. 

\subsubsection{Attacks against integrity} We classify six attacks in this category, namely, spam attack, message blocking, cloning attack, message modification attack, message insertion attack, and tampering attack. Note that the Spam attack can be classified in the category of attacks against availability.  An attack against integrity is based on the modification of a data exchanged between the 5G access points and the mobile users. However, the authentication and privacy preserving schemes for 4G and 5G cellular networks use mostly the hash functions for assuring integrity of transmitted data. The SHA-1 and MD5 algorithms are frequently used as hash functions, which can easily detect the attacks against integrity by verifying an incorrect hash value.\textbf{}

\subsubsection{Attacks against availability} We classify six attacks in this category, namely, First In First Out (FIFO) attack, redirection attack, physical attack, skimming attack, and free-riding attack. The goal of an attack against availability is to make a service as unavailable, e.g., the data routing service. By gathering entering time and exiting time intervals, the FIFO attack can be launched by a strong adversary. Gao et al. \cite{88} discuss the FIFO attack and propose a trajectory mix-zones graph model. The redirection attack is easily possible when an adversary gets the correct user entity information by increase its signal strength to redirect or by impersonating a base station in the 4G and 5G cellular networks. To protect the network from redirection attack, Saxena et al. \cite{116} and Li et al. \cite{121} proposed the same idea that uses a MAC to maintain the integrity of tracking area identity, while Yao et al. \cite{122} uses the local area identifier embedded with MAC. Therefore, the free-riding attack can cause a serious threat and reduces the system availability of D2D communication in the 4G and 5G cellular networks. By keeping a record of the current status of the user equipment and realize reception non-repudiation by key hint transmission, the proposed protocol by Zhang et al. \cite{117} can detect the free-riding attack. 

\subsubsection{Attacks against authentication} We classify ten attacks in this category, namely, password reuse attack, password stealing attack, dictionary attack, brute force attack, desynchronization attack, forgery attack, leak of verifier attack, partial-message collision attack, and stolen smart card attack. The goal of an attack against authentication is to disrupt the client-to-server authentication and the server-to-client authentication. The password reuse attack and password stealing attack disrupt the password-based authentication schemes, which the attacker pretends to be legitimate user and attempts to login on to the server by guessing different words as password from a dictionary. The stolen smart card attack and off-line guessing attack disrupt the smart-card-based remote user password authentication schemes, which if a user's smart card is stolen, the attacker can extract the stored information without knowing any passwords.
\begin{table}[!t]
	\centering
	\caption{Countermeasures used by the authentication and privacy preserving schemes for 4G and 5G cellular networks}
	\label{Table:Tab4}
	\setlength{\tabcolsep}{2pt}
		\vspace*{-\baselineskip}
		\renewcommand{\arraystretch}{1.5}
{\tiny \begin{tabular}{p{1.7in}|p{1.6in}} \hline 
\textbf{Countermeasures} & \textbf{Authentication and privacy preserving schemes that use the countermeasure} \\ \hline \hline
Secure hash function & \cite{57} , \cite{58}, \cite{59}, \cite{60}, \cite{61}, \cite{63}, \cite{65}, \cite{67}, \cite{68}, \cite{69}, \cite{71}, \cite{73}, \cite{75}, \cite{79}, \cite{80}, \cite{81}, \cite{82}, \cite{83}, \cite{84}, \cite{85}, \cite{86}, \cite{87}, \cite{91}, \cite{92}, \cite{93}, \cite{94}, \cite{95}, \cite{96}, \cite{97}, \cite{98}, \cite{99}, \cite{100}, \cite{101}, \cite{102}, \cite{103}, \cite{104}, \cite{105}, \cite{107}, \cite{108}, \cite{111}, \cite{112}, \cite{113}, \cite{114}, \cite{115}, \cite{117} \\ \hline 
Index-pseudonym & \cite{73} \\ \hline 
UMTS-AKA mechanism & \cite{53}, \cite{54} \\ \hline 
Message Authentication Code (MAC) & \cite{54}, \cite{60}, \cite{82}, \cite{91} \\ \hline 
Electronic Product Code (EPC) & \cite{56}, \cite{63} \\ \hline 
Intrusion Detection Message Exchange Format (IDMEF) and  Intrusion Detection Exchange Protocol (IDXP) & \cite{78} \\ \hline 
Digital certificate and signature & \cite{59}, \cite{117} \\ \hline 
Public Key Infrastructure (PKI) & \cite{59}, \cite{62}, \cite{76} , \cite{98}, \cite{99} \\ \hline 
Advanced Encryption Standard (AES) & \cite{58} \\ \hline 
APFS protocol  and  Digital signature standard (DSS) & \cite{60} \\ \hline 
Password & \cite{61}, \cite{67}, \cite{79}, \cite{80}, \cite{85}, \cite{86} \\ \hline 
Transport Layer Security (TLS) & \cite{61}, \cite{76} \\ \hline 
Trusted Platform Module (TPM) & \cite{61} \\ \hline 
Keyed-Hash Message Authentication Code (HMAC) & \cite{63}, \cite{75}, \cite{93}, \cite{116}, \cite{117} \\ \hline 
Pseudorandom Number Generator (PRNG) & \cite{64}, \cite{73}, \cite{108} \\ \hline 
Cyclic Redundancy Code (CRC-16) & \cite{64}, \cite{108} \\ \hline 
Homomorphic Encryption & \cite{114}, \cite{74} \\ \hline 
Paillier cryptosystem & \cite{114}, \cite{74} \\ \hline 
Forward security technique & \cite{65} \\ \hline 
Error Correction Codes (ECC) & \cite{99}, \cite{66} \\ \hline 
Anonymous ticket & \cite{77} \\ \hline 
Biometrics & \cite{67} \\ \hline 
Blind signature and  Rabin's public key cryptosystem & \cite{68} \\ \hline 
Elliptic Curve Diffie--Hellman protocol (ECDH) & \cite{105}, \cite{111}, \cite{69}, \cite{83} \\ \hline 
Bootstrapping Pseudonym (BP), Home Fast Pseudonym (HFP), and Visited Fast Pseudonym (VFP) & \cite{70} \\ \hline 
Advanced Identity Management (AIM) & \cite{71} \\ \hline 
Physically Unclonable Function (PUF) & \cite{72} \\ \hline 
Linear Feedback Shift Register (LFSR) & \cite{72} \\ \hline 
Personal Identification Number (PIN) & \cite{53} \\ \hline 
Semantic secure symmetric encryption & \cite{65} \\ \hline 
Smart cards & \cite{100}, \cite{101}, \cite{104}, \cite{111}, \cite{67} \\ \hline 
Proxy-signature scheme & \cite{81} \\ \hline 
Network domain security (NDS)/IP & \cite{83} \\ \hline 
Trusted Node Authentication (TNA) & \cite{84} \\ \hline 
Schnorr's signature scheme & \cite{87} \\ \hline 
Pseudo-Location Swapping (PLS) & \cite{89} \\ \hline 
Symmetric encryption & \cite{91}, \cite{96}, \cite{116} \\ \hline 
Hierarchical identity-based signature & \cite{92} \\ \hline 
Mobile vector network protocol & \cite{92} \\ \hline 
Hamming weight of vector & \cite{94} \\ \hline 
International Mobile Subscriber Identity (IMSI) & \cite{95}, \cite{115}, \cite{116} \\ \hline 
Radio Network Temporary Identities (RNTI) & \cite{95} \\ \hline 
Fuzzy extractor & \cite{101}, \cite{111} \\ \hline 
Certificate revocation & \cite{102} \\ \hline 
Group signatures with verifier local revocation & \cite{103} \\ \hline 
Group-based access authentication & \cite{106} \\ \hline 
Aggregate Message Authentication Codes AMAC & \cite{107}, \cite{113} \\ \hline 
Designated verifier proxy signature (DVPS) & \cite{115} \\ \hline 
\end{tabular}}
\end{table}

\begin{figure}[h]
\centering
\includegraphics[width=1\linewidth]{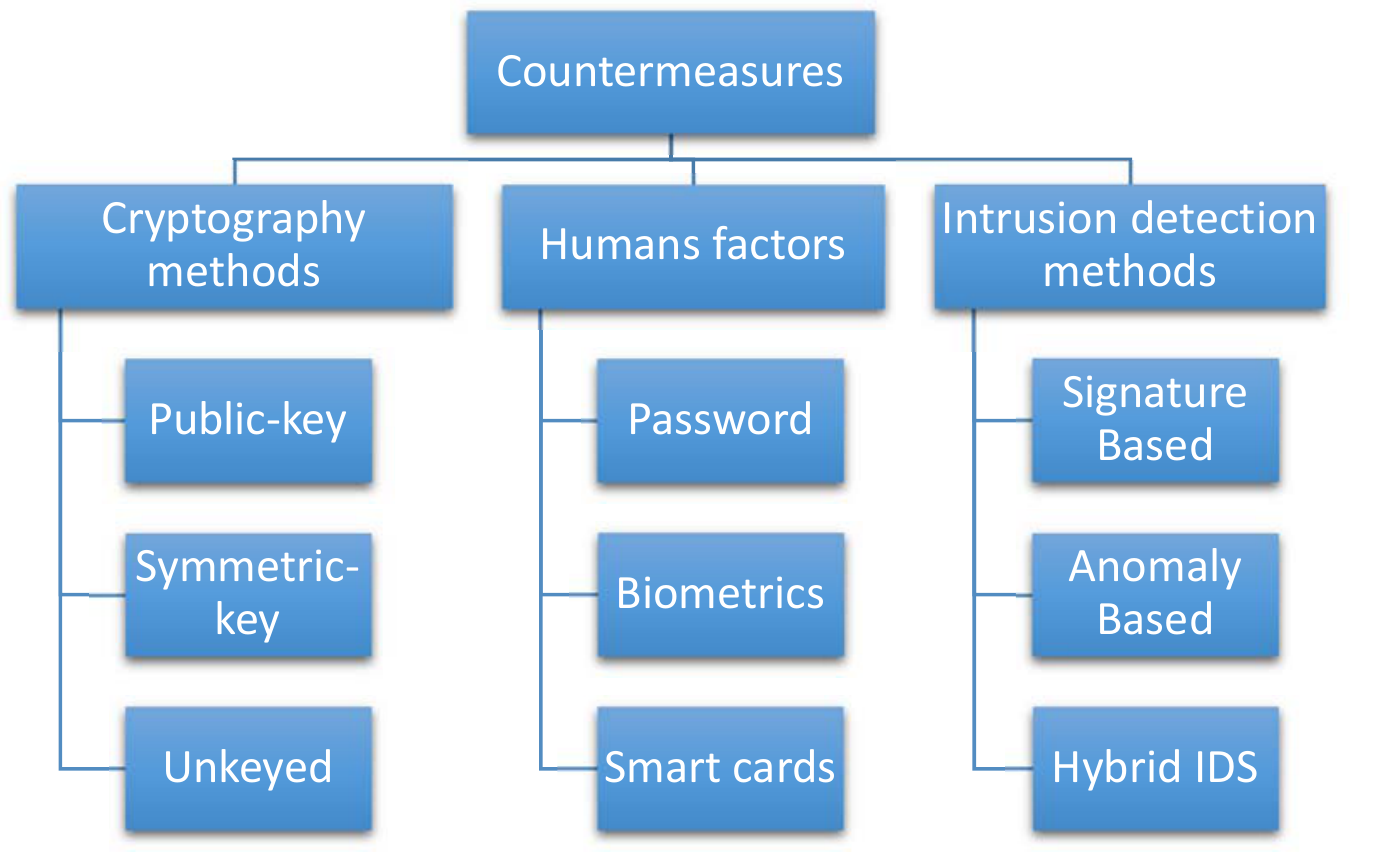}
\caption{Classification of countermeasures used by the authentication and privacy preserving schemes for 4G and 5G cellular networks}
\label{fig:Fig3_2}
\end{figure}
\begin{figure*}
\centering
\includegraphics[width=0.9\linewidth]{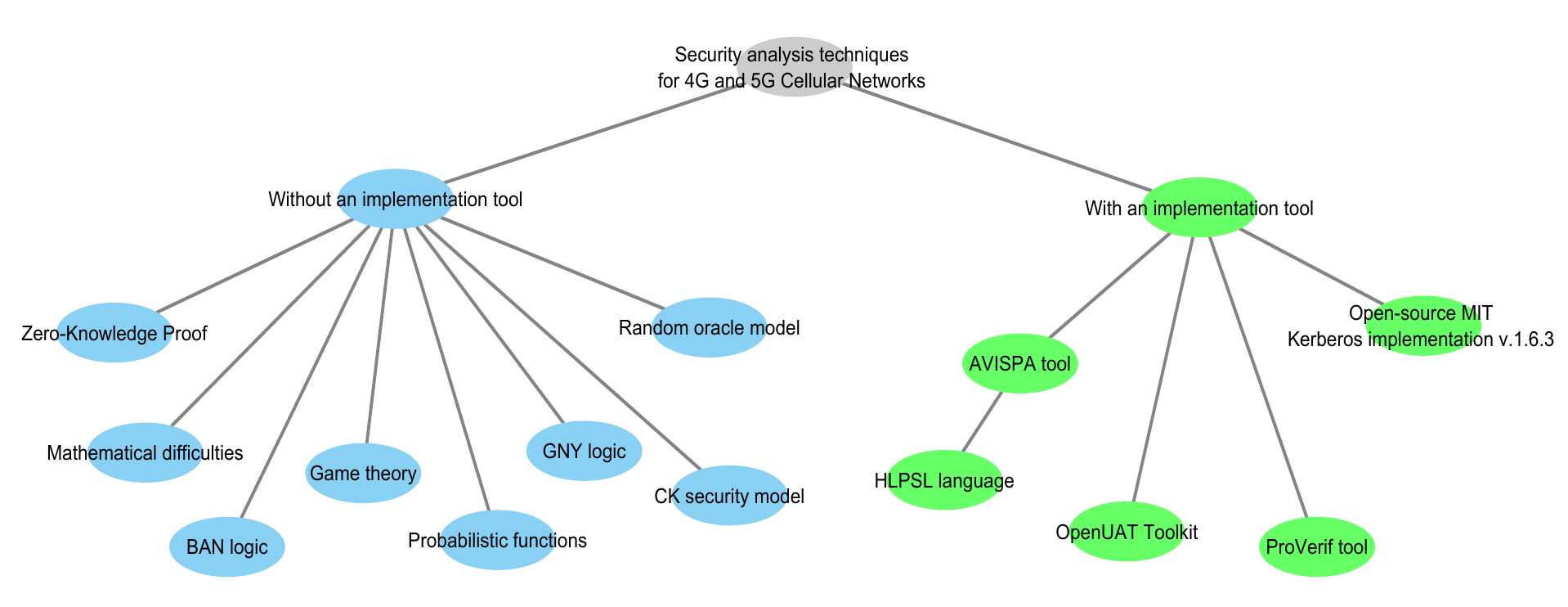}
\caption{Classification of security analysis techniques}
\label{fig:Fig4_1}
\end{figure*}
\subsection{Countermeasures}\label{sec:countermeasures}
In this subsection, we discuss the countermeasures used by the authentication and privacy preserving schemes for 4G and 5G cellular networks. Tab. \ref{Table:Tab4}presents all the countermeasures used by the authentication and privacy preserving schemes for 4G and 5G cellular networks. These countermeasures can be classified into three types of categories, including, cryptography methods, humans factors, and intrusion detection methods, as presented in Fig. \ref{fig:Fig3_2}.

\subsubsection{Cryptography methods}

Cryptographic methods are the most used by the authentication and privacy preserving schemes for 4G and 5G cellular networks, which can be classified into three types of categories, including, public-key cryptography, symmetric-key cryptography, and unkeyed cryptography. 

The schemes \cite{59}, \cite{62}, \cite{76}, \cite{98}, and \cite{99} use the public key infrastructure (PKI) \cite{146} in order to identify the genuine access point (AP) or base station (BS). Both schemes \cite{114}, \cite{74} use the Paillier cryptosystem \cite{204}, which is based on three algorithms, namely, \textit{generation of keys}, \textit{encryption}, and \textit{decryption}. The \textit{generation of keys} is based on two large, independent and random prime numbers: $p$ and $q$. Let $m$ be a message to be encrypted, the \textit{encryption }algorithm\textit{ }computes $c={(1+N)}^m\cdot r^N{\rm mod\ }N^2$ where $0\le m<N$, $r$ is a random integer $0<r<N$, and the public key $N=p\cdot q$. To find the clear text $m$, the \textit{decryption }algorithm\textit{ }computes $m=\frac{\left(c.r^{-N}{\rm mod}\ N^2\right)-1}{N}$. The scheme \cite{68} uses both Blind signature \cite{185} and Rabin's public key cryptosystem \cite{184}. The blind signature involves two entities, namely: 1) a signer and 2) a signature requester, in which the content of a message is disguised from its signature. Rabin's public key cryptosystem is characterized by its asymmetric computational cost and requires a large amount of computation effort. The Group signatures with verifier local revocation \cite{274} is used by the scheme \cite{103} in order to provide conditional anonymity. Furthermore, Boneh et al. \cite{273} proposed short group signatures due to group signatures based on Strong-RSA are too long for some applications. The digital signature standard (DSS) \cite{153} is used by the PT scheme \cite{60} in order to provide confidentiality and integrity to data exchanges after authentication as well as to simplify the key exchange protocol.

The symmetric encryption is used by four schemes, namely, \cite{91}, \cite{96}, \cite{116}, \cite{65}, in order to provide user anonymity. Specifically, Chen et al. \cite{91} use the Advanced Encryption Standard (AES) as the symmetric data encryption algorithm for mobile devices. Based on the idea that symmetric key algorithms faster than asymmetric key algorithms, Saxena et al. \cite{116} proposed an authentication protocol that is entirely based on the symmetric key cryptosystem for an IoT-enabled LTE network. Therefore, the question we ask here is: can the strategy of only using symmetric key techniques to achieve user anonymity is reliable? The improved privacy-preserving authentication scheme proposed recently by Wang et al. in \cite{96} can answer this question where he can proved that the strategy of only using symmetric-key techniques to achieve user anonymity is intrinsically infeasible. In addition, Lu et al. \cite{65} use semantic secure symmetric encryption in order to preserve the location privacy.

Hash functions are used almost in all the authentication and privacy preserving schemes in order to provide data integrity for the encrypted messages. We note that these schemes use three popular methods, namely, the Message Authentication Code (MAC) \cite{131}, the Keyed-Hash Message Authentication Code (HMAC) \cite{162}, and the Aggregate Message Authentication Codes (AMAC) \cite{287}. 

\subsubsection{Humans factors}

The humans' factors-based countermeasures are proposed to ensure authentication. The research community has proposed three factors, namely, 1) what you know (e.g., passwords, personal identification number (PIN)), 2) what you have (e.g., token, smart cards, passcodes, RFID), and 3) who are you (e.g., biometrics like fingerprints and iris scan, signature or voice). The methods based on what you know (e.g., passwords) might be divulged or forgotten, and the methods based on what you have (e.g., smart cards) might be shared, lost, or stolen. In contrast, the methods based on who are you (e.g., fingerprints or iris scans) have no such drawbacks. Note that these three factors can be used together or alone. 

\subsubsection{Intrusion detection methods}

Intrusion Detection systems (IDS) are the second stage of defense. In situations when an intruder has already managed to bypass all existing countermeasures and has already taken control of a legal entity of the network, an IDS must spot misbehavior fast enough in order to be efficient. There are a lot of new methods that have been proposed during the previous years for detecting intruders in 4G and 5G networks. In \cite{305} authors propose a novel IDS based on Bayesian Robust Principal Component Analysis (BRPCA). Based on the observation that network traffic variables are non-stationary and exhibit 24 h periodicity, the proposed anomaly detection approach represents network traffic as a sequence of traffic variable vectors. The method was evaluated against two synthetic datasets that represent a DOS and femtocell-based attack respectively. Trying to combat a similar attack, a virtual jamming attack, authors in \cite{306} proposed a novel hybrid NIDS based on Dempster-Shafer (DS) Theory of Evidence. The performance of the method, that combines a signature-based and an anomaly based IDS, was evaluated on an experimental IEEE 802.11 network testbed.  

In \cite{307} authors propose an adaptive intrusion detection system that uses a hidden Markov Model for detecting intrusions on small cell access point in a 5G wireless communication networks. Authors focused on the bandwidth spoofing attack. During this attack, the attacker tries to acquire the bandwidth that is going to be assigned from the BS to the SCA, thus blocking its communication. The method is proved to be capable of detecting and removing the intruder which is executing a bandwidth spoofing attack on the SCA (small cell access) in a 5G WCN.  In \cite{308} authors proposed an RNN-based  (Random Neural Network) approach for detecting of large scale Internet anomalies based on the analysis of captured network data. Authors we mostly interested in investigating application specific anomalies and conducted the evaluation of their proposed method on semi-synthetic data, derived from real traffic traces.  Relying on fuzzy logic principles, authors propose in \cite{309} a novel Intrusion Detection System.  The proposed IDS uses an Adaptive Neuro-Fuzzy Inference System and is created for 5G Wireless Communication Network (WCN). The proposed IDS is a fuzzy inference system integrated with neural networks taking advantage of the benefits of both systems \cite{310}.  Authors evaluated their method against DOS attacks, like the previous methods in \cite{305,306,307,308} using the KDD cup 99 dataset. In a scenario where malicious data packets coming from a 3G, 4G or Wi-Fi network that the vehicle use in order to communicate with surrounding vehicles, manage to enter into the in-vehicle CAN bus is investigated in \cite{311}. Authors in \cite{311} propose IDS that uses a deep neural network (DNN) in order to detect an attack after it has entered the CAN (controller area network). The proposed IDS provides a real-time response to the attack with good accuracy.  

Dealing with attacks in LTE networks, authors in \cite{312} propose a random packet inspection scheme. The proposed scheme has an inspection rate that can be dynamically adjusted based on the perceived intrusion period of the session. This way the IDS performs a deep packet inspection, which is necessary in order to reveal the presence of signatures or malicious codes, while on the same time being an efficient and quick way of inspection. This method provides an effective tool for balancing induced inspection cost with  detection latency in LTE core networks. In \cite{313} authors cope with intrusions in wireless sensor networks. The authors having identified the key aspects of such a network, e.g. Highly dynamic network conditions, limited bandwidth and transmission of sensitive data, propose TermID and test its efficiency using the Aegean wireless intrusion dataset version 2 \cite{314}. The proposed method achieves both low network footprint and user privacy. Taking in mind privacy along with security, authors in \cite{78} propose a location aware mobile IDS system. The proposed mIPS is a location-aware intrusion detection and prevention system with enhanced privacy handling.

Intelligence of intruders affects the effectiveness of IDS. This situation is investigated in \cite{316} where authors implemented two AI-enabled intrusion algorithms and evaluated the impact of intruder's intelligence on the intrusion detection capability of a WSN under various circumstances. Moving one step further, authors in \cite{315} review the area of Intrusion Response Systems (IRS). An IRS taking in mind the current situation on the network may choose the optimal response option. Based on the research of the authors, IRS cannot handle false alarms that are produced from the IDS and in the future a false alarm handler is an important component that must be integrated in every IDS/IRS.
\begin{table*}[!t]
	\centering
	\caption{Informal and formal security analysis techniques used in authentication and privacy preserving schemes for 4G and 5G Cellular Networks}
	\label{Table:Tab5}
	\setlength{\tabcolsep}{2pt}
		\vspace*{-\baselineskip}
		\renewcommand{\arraystretch}{1.5}
{\scriptsize \begin{tabular}{p{0.2in}|p{0.2in}|p{1in}|p{1.2in}|p{1.2in}|p{2in}|p{0.4in}} \hline 
\textbf{Ref.} & \textbf{Year} & \textbf{Tool} & \textbf{Authentication model to prove} & \textbf{Privacy model to prove} & \textbf{Main results} & \textbf{Implem.} \\ \hline \hline
 \cite{54} & 2007 & - Communicating Sequential Processes (CSP) \cite{132};\newline - Rank Functions; & - Mutual authentication\newline - Biometric authentication & - Data privacy & - Formalize the authentication and key establishment properties of the IDM3G protocol as trace specifications. & No \\ \hline 
 \cite{60} & 2008 & - Zero-Knowledge Proof \cite{154} & - User authentication & - Mutual anonymity & - Analyze the anonymity degree of the PT protocol. & No \\ \hline 
 \cite{62} & 2009 & - Strand spaces model \cite{161} & - Authentication and Key Agreement & - Confidentiality & - Analyze security performance of the authentication and key agreement protocol. & No \\ \hline 
 \cite{67} & 2009 & - GNY logic \cite{183} & - Three-factor authentication\newline - Remote user authentication & - Privacy of the biometric data & - Analyze the completeness of a cryptographic protocol. & No \\ \hline 
 \cite{69} & 2009 & - CK security model \cite{188}\newline  & - Mutual authentication and key agreement & - N/A & - Prove that the NAKE protocol is probably secure. & No \\ \hline 
 \cite{70} & 2010 & - Network Address Identifier (NAI) format \cite{191} & - Fast re-authentication & - Identity privacy & - Test the privacy solution behavior. & No \\ \hline 
 \cite{71} & 2010 & - AVISPA tool \cite{193};\newline - HLPSL language \cite{193}; & - Mutual authentication & - Identity privacy & - Prove the efficiency of the identity management mechanism. & Yes \\ \hline 
 \cite{77} & 2011 & - Open-source MIT Kerberos implementation v.1.6.3 \cite{209} & - Cross-realm authentication & - Anonymity;\newline - Service access untraceability; & - Evaluate the performance of the enhanced Kerberos protocol. & Yes \\ \hline 
 \cite{81} & 2012 & - AVISPA tool \cite{193};\newline - HLPSL language \cite{193}; & - Handover authentication & - N/A & - Show that the scheme can work correctly to achieve robust security properties. & Yes \\ \hline 
 \cite{82} & 2012 & - AVISPA tool \cite{193};\newline - HLPSL language \cite{193}; & - Handover authentication & - Identity privacy & - Ensure the security of the handover authentication scheme. & Yes \\ \hline 
 \cite{86} & 2012 & - ProVerif \cite{235} & - Identity based authentication & - N/A & - Guarantee the necessary security features claimed by the oPass protocol. & Yes \\ \hline 
 \cite{87} & 2012 & - Random oracle model \cite{239} & - Authentication and key agreement & - N/A & - Show that there is an adversary A can construct an algorithm to solve the CDH problem or the k-CAA problem separately. & No \\ \hline 
 \cite{91} & 2013 & - Game theory \cite{243}\newline  & - Authentication and key agreement & - N/A & - Prove the security of the bipartite protocol by designing a game that turns a CDH instance into the protocol. & No \\ \hline 
 \cite{93} & 2013 & - OpenUAT \cite{248} & - Multichannel authentication & - N/A & - Implement some intuitive authentication methods in a common library based. & Yes \\ \hline 
 \cite{94} & 2013 & - Probabilistic functions \cite{250} & - RFID authentication & - N/A & - Define formally security models for the LCMQ authentication system. & No \\ \hline 
 \cite{103} & 2015 & - ProVerif \cite{235} & - Mutual authentication & - Location privacy & - Verify the system in $\pi -Calculus$ with ProVerif. & Yes \\ \hline 
 \cite{104} & 2015 & - ProVerif \cite{235}\newline - Game theory \cite{243} & - Remote user authentication & - Anonymity & - Verify the resistance against known attacks. & Yes \\ \hline 
 \cite{105} & 2015 & - Bellare--Rogaway \cite{281} & - Roaming authentication & - Anonymity & - Prove the security of scheme under Elliptic Curve Diffie--Hellman (ECDH) assumption. & No \\ \hline 
 \cite{110} & 2015 & - GNY logic \cite{183} & - RFID mutual authentication & - N/A & - Prove the correctness of the LRMAPC protocol. & No \\ \hline 
\cite{111} & 2015 & - BAN logic \cite{290} & - Biometrics-based authentication & - Anonymity & - Demonstrate that the scheme is valid and practical. & No \\ \hline 
 \cite{113} & 2016 & - ProVerif \cite{235} & - Group authentication; & - Anonymity;\newline - Unlinkability;\newline - Traceability; & - Verify the secrecy of the real identity. & Yes \\ \hline 
\cite{114} & 2016 & - Mathematical difficulties & - Anonymous authentication & - Location privacy & - Achieve security and privacy using discrete logarithm and computational Diffie-Hellman problems. & No \\ \hline 
\cite{120} & 2016 & - AVISPA tool \cite{193} & - Mutual authentication with key agreement & - Location privacy & - Verify the protocol security against insider attacks and outsider attacks. & Yes \\ \hline 
\cite{123} & 2016 & - AVISPA tool \cite{193} & - Handover authentication & - Anonymity;\newline - Unlinkability;\newline - Traceability;\newline - Non-frameability; & - Show that \textit{Nframe} can maintain the security requirements in frequent handover authentication semantics. & Yes \\ \hline 
\end{tabular}}
\end{table*}

\section{Informal and formal security analysis techniques}\label{sec:informal-and-formal-security-analysis-techniques}
Researchers in the Security and Privacy fields use the formal and informal techniques to analyze, prove, and verify the reliability of their proposed security scheme, and especially for schemes that are based on cryptography as a tool for achieving the authentication and privacy. Therefore, we classify these techniques on two classes, including, 1) \textit{Without an implementation tool} and 2) \textit{With an implementation tool}, as presented in Fig. \ref{fig:Fig4_1}. In addition, Tab. \ref{Table:Tab5}. summarizes the informal and formal security analysis techniques used in authentication and privacy preserving schemes for 4G and 5G Cellular Networks.

For the first class, we classify in '\textit{without an implementation tool}' eight techniques, including, Zero-Knowledge Proof \cite{154}, Mathematical difficulties, GNY logic \cite{183}, CK security model \cite{188}, Random oracle model \cite{239}, Game theory \cite{243}, Probabilistic functions \cite{250}, and BAN logic \cite{290}. To analyze the completeness of a cryptographic protocol, both schemes \cite{110} and \cite{67} use the GNY logic \cite{183}. The scheme \cite{87} use random oracle model \cite{239} to show that there is an adversary A can construct an algorithm to solve the CDH problem or the k-CAA problem separately. The scheme \cite{111} uses the BAN logic \cite{290} to demonstrate that the scheme is valid and practical. The mathematical difficulties is used by the scheme \cite{114} to achieve security and privacy using discrete logarithm and computational Diffie-Hellman problems. Furthermore, the game theory \cite{243} is used by the scheme \cite{91} to prove the security of the bipartite protocol by designing a game that turns a CDH instance into the protocol. According to Manshaei et al. \cite{243}, the game approach is related to the security problem to be solved, e.g., the \textit{stackelberg game} for \textit{Jamming/Eavesdropping}, the \textit{static security cost game} for \textit{Interdependent Security}, and the \textit{static non-zerosum game} for \textit{Vendor Patch Management}.

For the second class, we classify in 'with an implementation tool' four techniques, including, AVISPA tool \cite{193}, Open-source MIT Kerberos implementation v.1.6.3 \cite{209}, OpenUAT \cite{248}, and ProVerif \cite{235}. The Open-source MIT Kerberos \cite{209} is used especially for evaluate the performance of the enhanced Kerberos protocol such as the scheme \cite{77}. The OpenUAT \cite{248} is used by the scheme \cite{93} to implement some intuitive authentication methods in a common library. To verify the secrecy of the real identity and the resistance against known attacks, four schemes \cite{113}, \cite{104}, \cite{103}, and \cite{103} use the ProVerif \cite{235}, which is an automatic cryptographic protocol verifier, in the formal model, called Dolev-Yao model. Specifically, the ProVerif takes as input a model of the protocol in an extension of the pi calculus with cryptography. For more details about the ProVerif, we refer the reader to the work of Blanchet in \cite{303}. Therefore, five schemes \cite{120}, \cite{123}, \cite{81}, \cite{82}, and \cite{71} use the AVISPA tool \cite{193} based on the HLPSL language \cite{193} to verify the security of these schemes against insider attacks and outsider attacks.
\begin{table*}[!t]
	\centering
	\caption{Authentication models achieved by security schemes for 4G and 5G Cellular Networks}
	\label{Table:Tab6}
	\setlength{\tabcolsep}{2pt}
		\vspace*{-\baselineskip}
		\renewcommand{\arraystretch}{1.5}
		{\tiny \begin{tabular}{p{1.5in}|p{0.4in}|p{0.5in}|p{0.4in}|p{0.4in}|p{0.4in}|p{0.4in}|p{0.4in}|p{0.5in}|p{0.4in}|p{0.4in}|p{0.4in}} \hline 
		 & \multicolumn{11}{|p{4.6in}}{\textbf{  Authentication models}} \\ \hline \hline
		\textbf{Schemes} & \textbf{Mutual authen.} & \textbf{Identity-based authen.} & \textbf{Remote user authen.} & \textbf{Key agreement} & \textbf{RFID authen.} & \textbf{Fast re-authen.} & \textbf{Three-factor authen.} & \textbf{Password-based authen.} & \textbf{Deniable authen.} & \textbf{Biometric authen.} & \textbf{Handover authen.} \\ \hline \hline 
		\cite{53} \cite{54} \cite{58} \cite{59} \cite{65} \cite{68} \cite{69} \cite{71} \cite{75} \cite{89} \cite{95} \cite{97} \cite{98} \cite{100} \cite{102} \cite{103} \cite{115} \cite{116} \cite{118} \cite{120} & \cellcolor{black!25} &  &  &  &  &  &  &  &  &  &  \\ \hline 
		\cite{54} \cite{67} \cite{111}\newline  &  &  &  &  &  &  &  &  &  & \cellcolor{black!25} &  \\ \hline 
		\cite{55} \cite{56} \cite{63} \cite{64} \cite{66} \cite{72} \cite{73} \cite{94} \cite{99} \cite{110} &  &  &  &  & \cellcolor{black!25} &  &  &  &  &  &  \\ \hline 
		\cite{57} \cite{61}\newline  &  &  &  &  &  &  &  &  & \cellcolor{black!25} &  &  \\ \hline 
		\cite{62} \cite{68} \cite{69} \cite{82} \cite{83} \cite{85} \cite{87} \cite{91} \cite{97} \cite{100} \cite{102} \cite{107} \cite{115} \cite{120} &  &  &  & \cellcolor{black!25} &  &  &  &  &  &  &  \\ \hline 
		\cite{67} \cite{101}\newline  &  &  &  &  &  &  & \cellcolor{black!25} &  &  &  &  \\ \hline 
		\cite{67} \cite{80} \cite{101} \cite{104} &  &  & \cellcolor{black!25} &  &  &  &  &  &  &  &  \\ \hline 
		\cite{70}\newline  &  &  &  &  &  & \cellcolor{black!25} &  &  &  &  &  \\ \hline 
		\cite{79} \cite{96} \cite{112}\newline  &  &  &  &  &  &  &  & \cellcolor{black!25} &  &  &  \\ \hline 
		\cite{81} \cite{82} \cite{83} \cite{106} \cite{107} \cite{123} &  &  &  &  &  &  &  &  &  &  & \cellcolor{black!25} \\ \hline 
		\cite{85} \cite{86} \cite{92}\newline  &  & \cellcolor{black!25} &  &  &  &  &  &  &  &  &  \\ \hline 
		\end{tabular}}
\end{table*}
\begin{table*}[!t]
	\centering
	\caption{Privacy models achieved by security schemes for 4G and 5G Cellular Networks}
	\label{Table:Tab7}
	\setlength{\tabcolsep}{2pt}
		\vspace*{-\baselineskip}
		\renewcommand{\arraystretch}{1.5}
	{\tiny 	\begin{tabular}{p{1.5in}|p{0.4in}|p{0.4in}|p{0.4in}|p{0.4in}|p{0.4in}|p{0.4in}|p{0.4in}|p{0.4in}|p{0.4in}|p{0.6in}} \hline \hline
		\textbf{} & \multicolumn{10}{|p{4.7in}}{\textbf{Privacy models}} \\ \hline 
		\textbf{Schemes} & \textbf{Identity privacy} & \textbf{Location privacy} & \textbf{Anonymity} & \textbf{RFID privacy} & \textbf{Untraceability} & \textbf{Non-frameability} & \textbf{Traceability} & \textbf{Conditional privacy} & \textbf{Forward privacy} & \textbf{Privacy preserving data aggregation} \\ \hline \hline
		\cite{113} \cite{123}\newline  &  &  &  &  &  &  & \cellcolor{black!25} &  &  &  \\ \hline 
		\cite{117} \cite{118}\newline  &  &  &  &  &  &  &  & \cellcolor{black!25} &  &  \\ \hline 
		\cite{53} \cite{59} \cite{70} \cite{71} \cite{74} \cite{76} \cite{82} \cite{115} \cite{79} & \cellcolor{black!25} &  &  &  &  &  &  &  &  &  \\ \hline 
		\cite{123}\newline  &  &  &  &  &  & \cellcolor{black!25} &  &  &  &  \\ \hline 
		\cite{56} \cite{55} \cite{73} \cite{96} \cite{99} \cite{103} \cite{104} \cite{105} \cite{106} \cite{111} \cite{113} \cite{116} \cite{118} \cite{123} \cite{77} &  &  & \cellcolor{black!25} &  &  &  &  &  &  &  \\ \hline 
		\cite{63} \cite{64} \cite{66} &  &  &  & \cellcolor{black!25} &  &  &  &  &  &  \\ \hline 
		\cite{65} \cite{73} \cite{75} \cite{88} \cite{89} \cite{99} \cite{103} \cite{114} \cite{120} &  & \cellcolor{black!25} &  &  &  &  &  &  &  &  \\ \hline 
		\cite{70} \cite{77} \cite{105} \cite{116} &  &  &  &  & \cellcolor{black!25} &  &  &  &  &  \\ \hline 
		\cite{116}\newline  &  &  &  &  &  &  &  &  & \cellcolor{black!25} &  \\ \hline 
		\cite{102}\newline  &  &  &  &  &  &  &  &  &  & \cellcolor{black!25} \\ \hline 
		\end{tabular}}
\end{table*}
\begin{table}[!t]
	\centering
	\caption{Notations used in comparison of computational cost and communication overhead}
	\label{Table:Tab9}
	\setlength{\tabcolsep}{2pt}
		\vspace*{-\baselineskip}
		\renewcommand{\arraystretch}{1.5}
{\scriptsize
\begin{tabular}{p{0.4in}|p{2.8in}} \hline
Notation & Definition \\ \hline \hline
$T_E\ $& The time complexity for exponentiation \\ $T_{SE}$& The time complexity for small-exponent exponentiation \\ $T_H$& The time complexity for hash function \\ $T_S$& The time complexity for symmetric encryption/decryption \\ $T_M$& The computation cost of multiplication operation\\ $T_{ECC}$& The time complexity for ECC-based scalar multiplication\\ $T_{COM}~$& The time to upload the encrypted traffic using 5G communication links\\ $T_L$& Lagrange component time\\ $e~$& The cost between the MTC device and the eNB\\ $\eta ~$& The cost between mobility management entities\\ $n$& The  number of MTC device\\ $m$& The number of groups\\ \hline 
\end{tabular}}
\end{table}
\begin{table*}
	\centering
	\caption{Summary of authentication and privacy preserving schemes for 4G and 5G Cellular Networks}
	\label{Table:Tab8}
	\setlength{\tabcolsep}{2pt}
		\vspace*{-\baselineskip}
		\renewcommand{\arraystretch}{1.5}
{\scriptsize
See Tab. \ref{Table:Tab9} for the notations used.
\begin{tabular}{p{0.37in}|p{1.3in}|p{0.7in}|p{0.7in}|p{2.7in}|p{1in}} \hline\hline
\textbf{Scheme} & \textbf{Network model} & \textbf{Auth. model} & \textbf{Privacy model} & \textbf{Performances (+) and limitations (-)} & \textbf{Complexity} \\ \hline 
Saxena et al. \cite{116} & - LTE cellular system with four entities, including, user equipment (UE), mobility management entity (MME), home service server (HSS), and radio access point& - Mutual authentication & - Untraceability;\newline - Forward privacy;\newline - Anonymity; & + Secure against replay attack, man-in-the-middle attack, redirection attack, impersonation attack, and message modification attack;\newline + Provide the untraceability, forward privacy, and anonymity;\newline + IoT-enabled LTE network;\newline + Reduce bandwidth consumption during authentication;\newline - The scalability is not considered compared to three schemes \cite{64}, \cite{59}, and \cite{60}. &Bandwidth consumption:
- Between UE and MME = 697 bits;\newline
- Between MME and HSS = 886 bits;
\\ \hline 
Wang et al. \cite{96} & - Roaming service in mobile networks& - Password-based authentication & - User anonymity & + Can achieve user anonymity;\newline + Can withstand offline password guessing attack even if the victim's smart card is lost;\newline + Efficient in term of computation cost on user side compared to fives schemes Li et al. \cite{252}, Isawa-Morii \cite{253}, He et al. \cite{254}, Zhou-Xu \cite{255}, and Xu et al. \cite{256};\newline - The proposed scheme needs to be evaluated in term of communication overhead;\newline - The handover delays are not measured; & - Computation cost on user side : $1T_{SE}\ +\ 4T_H\ +\ 1T_S$\\ \hline 
Cao et al.  \cite{106} & - Machine Type Communication (MTC) in LTE-A networks with three entities, including, the MTC device domain, the 3GPP network domain, and the MTC application domain& - Group-based handover authentication & - Anonymity & + Resistance to replay attack, eavesdropping attack, masquerade attack, and man-in-the-middle attack;\newline + Provide the security key derivation and anonymity;\newline - The scheme is not proven using the formal security analysis techniques. & - Signaling cost : $3n+5$ ;\newline
- Communication cost : $3en+4+\eta $
\\ \hline 
He and Wang \cite{111} & - Multiserver environment& - Biometrics-based authentication & - Anonymity & + Provide mutual authentication;\newline + Provide perfect forward secrecy;\newline + Suitable for the multiserver environment;\newline + Resistance to attacks, including, replay attack, stolen verifier attack, user impersonation attack, server spoofing attack, modification attack, and man-in-the-middle attack;\newline - The desynchronization is not considered. & - Computation cost on user side : $3T_M+7T_H$\\ \hline 
Fu et al. \cite{113} & - Machine-type communication (MTC) model in LTE advanced networks& - Group authentication; & - Anonymity;\newline - Unlinkability;\newline - Traceability; & + Provide robust privacy preserving including user anonymity, unlinkability, and traceability;\newline + Guarantee mutual authentication and congestion avoidance;\newline + Secure against four attacks, including, replay attack, impersonation attack, man-in-the-middle attack, and DoS attack;\newline - The desynchronization attack is not considered. & - Computation overhead: $(8n+6\ m)\ T_H+(3n+m)\ T_M$ ;\newline
- Signaling overhead: $n+6m$\\ \hline 
Mahmoud et al. \cite{114} & - LTE network-based advanced metering infrastructure (AMI)& - Anonymous authentication & - Location privacy & + Secure against impersonation attack, DoS attack, replay attack, and man-in-the-Middle attack;\newline + Verify the authenticity and integrity of the aggregated bids;\newline + Can achieve the privacy requirements with almost negligible performance degradation;\newline - The proposed scheme is not compared with other related schemes. &- The aggregated signature needs only 56 bytes\\ \hline 
Ramadan et al. \cite{115} &- LTE cellular system with four entities, including, user equipment, mobility management entity, home service server, and radio access point& - Mutual authentication and key agreement & - Identity privacy & + Secure against pro\eject ling attack, false base station attack, and replay attack;\newline + Provide the security architecture with flexibility and reliability;\newline + Provide forward/backward secrecy;\newline - The man-in-the-middle attack is not considered compared to the scheme \cite{94}; &- Computation cost on user side : $4T_{ECC}\ +\ 2T_H+\ 2T_P+\ T_M$\\ \hline 
Hashem Eiza et al. \cite{118} &-Multi-tier 5G enabled vehicular network with six entites, including, HetNets, D2D communications, a cloud platform, department of motor vehicles (DMV), trust authority (TA), and law enforcement agency (LEA)& - Mutual authentication & - Conditional anonymity ; \newline - Traceability of misbehaving participants& + Achieve the conditional anonymity and privacy;\newline + Resistant to traffic analysis attack, Sybil attack, eavesdropping attack, and fabrication attack;\newline + 5G enabled vehicular networks;\newline - The desynchronization attack is not considered. &- Computation cost on user side : $6T_H\ +\ 2T_S$; \newline - $T_{COM}$=13.3s\\ \hline 
Hamandi et al. \cite{120} & - LTE wireless network& - Mutual authentication with key agreement & - Location privacy & + Secure against replay attacks;\newline + Minimizes the use of both symmetric and asymmetric key encryption due to its excessive overhead;\newline - The untraceability and forward privacy are not considered.  & Signaling overhead: \newline
- Case with global random identity = 128 bits;
\\ \hline 
Li et al. \cite{121} &- Machine-type communication (MTC) model in LTE advanced network& - Group-based authentication & - N/A & + Secure against replay attack, redirection attack, man-in-the-middle attack, DoS attack, and impersonation attack;\newline + Can reduce the communication overhead and alleviates the burden between machine type communication devices;\newline - The known-key secrecy and the perfect forward secrecy are not considered compared to the scheme \cite{101}. &- Computation cost: $(T_L+T_H+2T_M)n$\\ \hline
Zhang et al. \cite{105} & - Roaming services in global mobility network & - Roaming authentication & - Anonymity & + Preserve the non-repudiation, user anonymity, and untraceability;\newline + Provide the perfect forward secrecy;\newline + Prevention of impersonation attacks;\newline - The DoS attack is not considered; & - Computation cost: $4T_M+{4T}_H+10T_S$ \\ \hline 
Fu et al. \cite{123} &- LTE/LTE-A network with the public switch telephone network& - Handover authentication & - Anonymity;\newline - Unlinkability;\newline - Traceability;\newline - Non-frameability; & + Protection against Man-in-the-Middle attack, DoS attack, and replay attack;\newline + Efficient in terms of computation cost and communication overhead compared to three schemes, namely, HALP scheme \cite{295}, Pair-Hand scheme \cite{296}, and UHAEN scheme \cite{297};\newline - The perfect forward and backward secrecy are not considered compared to the scheme \cite{81}. &- Computation cost on user side : $\ 5T_M$\\ \hline 
\end{tabular}\\}
\end{table*}
\begin{figure*}
\centering
\includegraphics[width=0.9\linewidth]{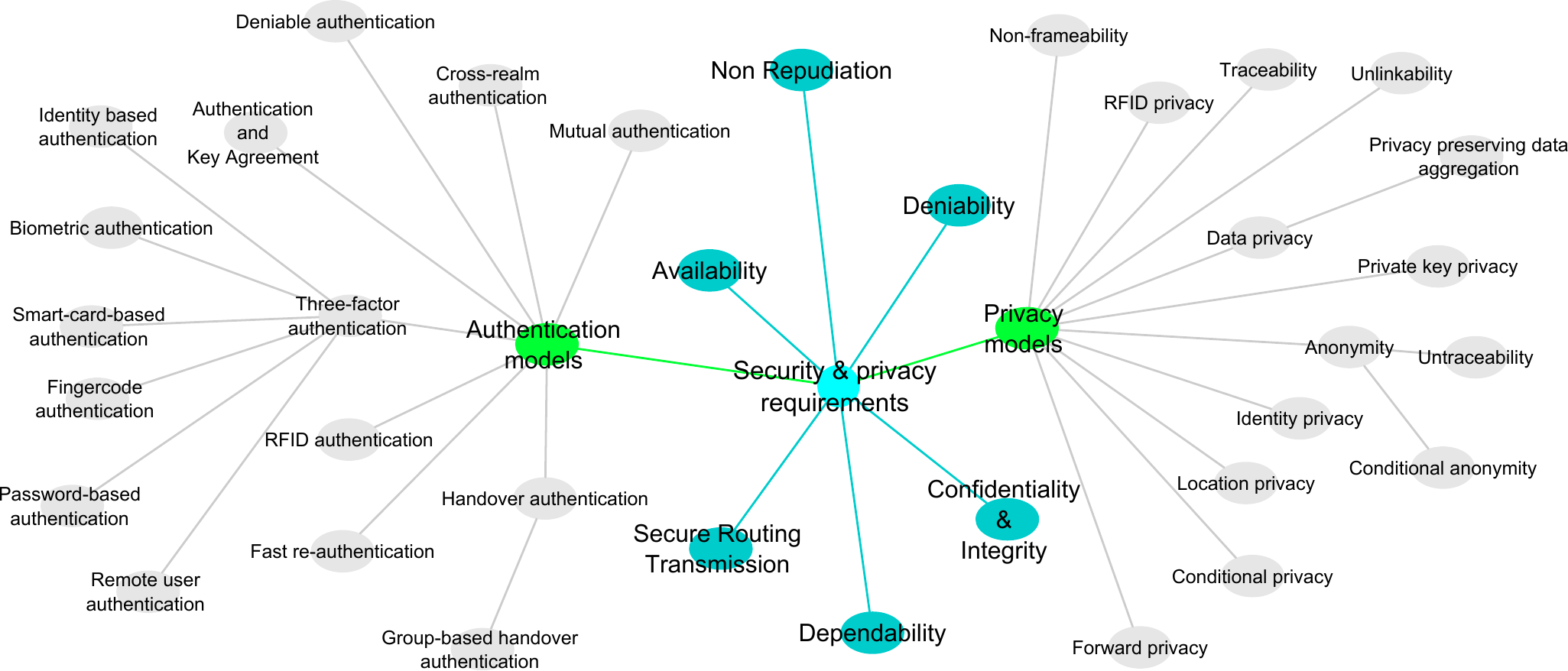}
\caption{Categorization of authentication and privacy models}
\label{fig:Fig6_1}
\end{figure*}
\section{Authentication and privacy preserving schemes for 4G and 5G Cellular Networks}\label{sec:authentication-and-privacy-preserving-schemes-for-4g-and-5g-cellular-networks}
\begin{figure}
\centering
\includegraphics[width=0.8\linewidth]{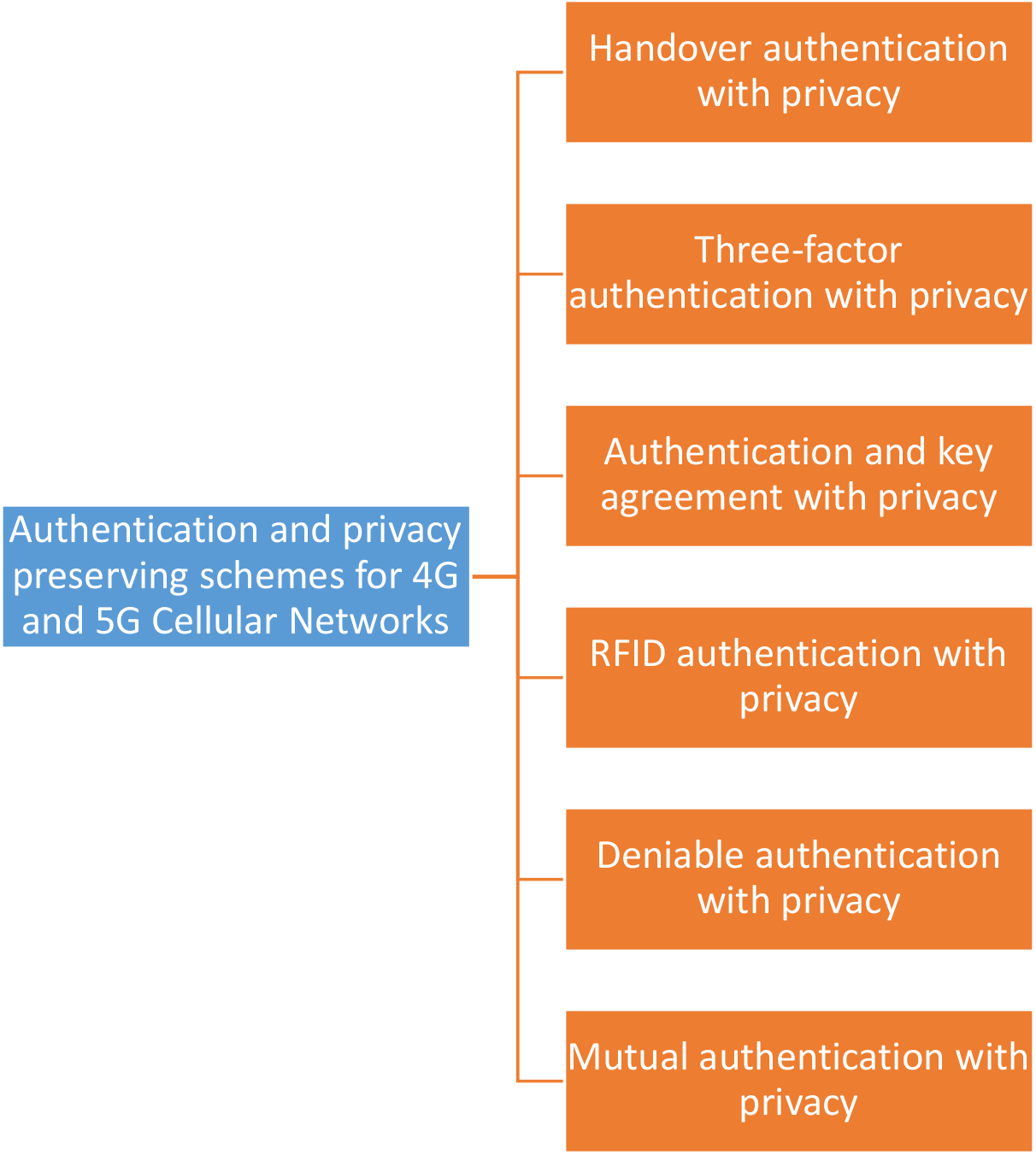}
\caption{Classification of authentication and privacy preserving schemes for 4G and 5G Cellular Networks}
\label{fig:Fig6_2}
\end{figure}

In this section, we will discuss the comparison of authentication and privacy preserving schemes for 4G and 5G Cellular Networks in term of authentication and privacy models. After reviewing around 50 papers published between 2005 and 2017, which are indexed in Scopus and Web of Science, we categorized the authentication and privacy models, as presented in Fig. \ref{fig:Fig6_1}, Tab\ref{Table:Tab6}, and Tab \ref{Table:Tab7}. Based on this categorization, we classify the schemes in seven types (as presented in Fig. \ref{fig:Fig6_2}), including, 1) \textit{Handover authentication with privacy}, 2) \textit{Mutual authentication with privacy}, 3) \textit{RFID authentication with privacy}, 4) \textit{Deniable authentication with privacy}, 5) \textit{Authentication with mutual anonymity}, 6) \textit{Authentication and key agreement with privacy}, and 7) \textit{Three-factor authentication with privacy}. Tab. \ref{Table:Tab8} summarizes the authentication and privacy preserving schemes for 4G and 5G Cellular Networks.

\subsection{Handover authentication with privacy}

Based on the cryptographic primitives, the existing handover authentication schemes for LTE wireless networks can be classified into three categories, including, 1) Symmetrical key-based scheme, 2) Public key-based scheme, and 3) Hybrid scheme. In LTE wireless networks, there are two types of base stations, namely, Home eNodeB (HeNB) and eNodeB (eNB). According to Cao et al. \cite{81}, the 3GPP project suggested handover from an eNB/HeNB to a new eNB/HeNB cannot achieve backward security in handover procedures. Specifically, the authors proposed a handover authentication scheme for the mobility scenarios in the LTE networks. Based on the idea of proxy signature, the scheme \cite{81} provide several security features, including, perfect forward and backward secrecy. In addition, the scheme \cite{81} is efficient in terms of computational cost and communication overhead compared with the  handover scheme in \cite{218}, but the identity privacy is not considered. Similar to the scheme \cite{81}, Cao et al. \cite{83} proposed a handover authentication scheme to fit in with all of the mobility scenarios in the LTE networks. The scheme can provide strong security guarantees including perfect forward secrecy, master key forward secrecy, and user anonymity. The scheme \cite{83} is efficient in terms of computational cost, communication cost, and storage cost. As a matter of fact, these both two schemes \cite{81} \cite{83} do not consider the identity and location privacy. To solve this problem, the idea of Gao et al. \cite{88} can be applied with both schemes \cite{81} and \cite{83}.

IEEE 802.16m is proposed as an advanced air interface to meet the requirements of the fourth generation (4G) systems. To preserves the identity privacy for IEEE 802.16m network, Fu et al. (2012) \cite{82} proposed a privacy-preserving fast handover authentication scheme based on the pseudonym. Based on the 3-way handshake procedure, the scheme \cite{82} can achieve the following research objectives, including, 1) Fast handover, 2) Mutual authentication and key agreement, and 3) Privacy preservation. In addition, the scheme \cite{82} is efficient in terms of computation and communication overhead compared with Fu et al. scheme \cite{229}. The scheme \cite{82} is does not consider k-anonymity, which is a privacy protection scheme with the context of location privacy. The following question is: Is it necessary to apply the k-anonymity improve user privacy in future 5G networks? Niu et al. \cite{89} show us that we need to generate and select the dummy users who can contribute to improving users privacy. As an additional benefit, the users can improve their location privacy significantly by applying the idea of pseudo-location swapping \cite{89}. 

To provide the security key derivation and anonymity for all of the mobility scenarios in LTE-A networks, Cao et al. \cite{106} proposed a group-based anonymity handover protocol, named NAHAP. The NAHAP protocol is efficient in terms of the signaling cost, the communication cost and the computational cost compared with the LTE-A handover mechanism. Similar to  NAHAP scheme, the same authors proposed another uniform group-based handover authentication protocol, named UGHA, which is efficient in term of computational cost compared with the scheme \cite{285}. Using software-defined networking, Duan and Wang \cite{109} proposed an authentication handover scheme with privacy protection in 5G heterogeneous network communications. Recently, Fu et al. \cite{113} proposed a novel group authentication protocol with privacy-preserving to provide unlinkability and traceability in 4G/5 communications. The scheme \cite{113} is efficient in terms of the signaling overhead and computation overhead compared to two schemes, including, Cao's scheme \cite{291} and SE-AKA \cite{292}. To fit in with all of the mobility scenarios in the LTA/LTA-A networks, Fu et al. \cite{123} proposed a privacy-preserving with non-frameability authentication protocol, called Nframe. To guarantee users' privacy, unlinkability and traceability, the Nframe protocol uses a pseudonym-based scheme. To achieve a simple authentication process without a complex key management and minimize message exchange time, the Nframe protocol uses pairing-free identity-based cryptography. In addition, the Nframe protocol is efficient in terms of computation cost and communication overhead compared to three schemes, namely, HALP scheme \cite{295}, Pair-Hand scheme \cite{296}, and UHAEN scheme \cite{297}, but the perfect forward and backward secrecy are not considered compared to the scheme \cite{81}. For more details in the field of handover authentication protocols using identity-based public key cryptography, we refer the reader to the recent work of He in \cite{304}.
\begin{figure}
\centering
\includegraphics[width=0.6\linewidth]{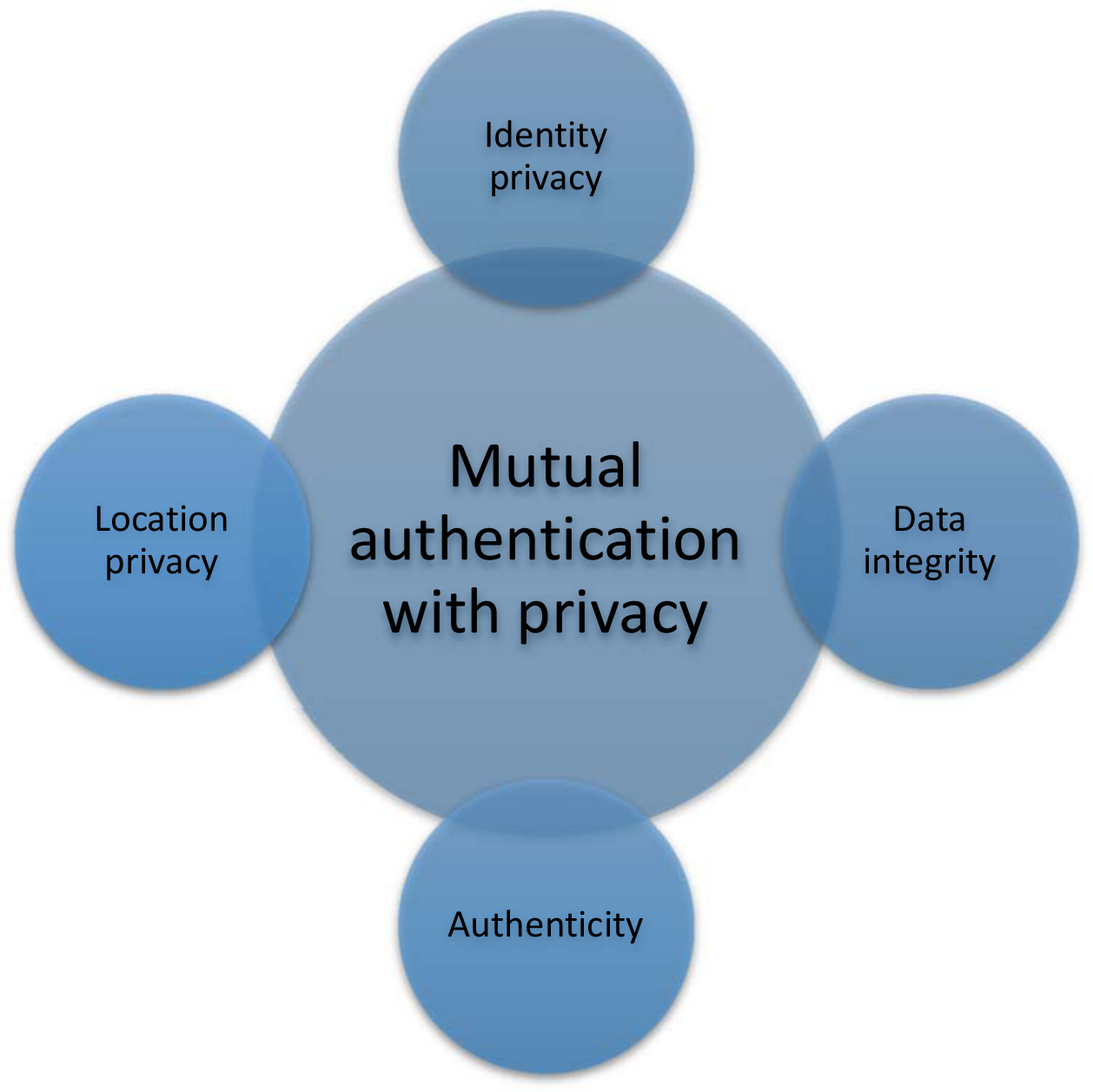}
\caption{Classification of mutual authentication with privacy schemes}
\label{fig:Fig6_4}
\end{figure}
\subsection{Mutual authentication with privacy}

To achieve the mutual authentication with privacy, the proposed security schemes for 4G/5G networks need to preserve the \textit{Location privacy}, \textit{Identity privacy}, \textit{Data integrity}, and \textit{Authenticity}, as shown in Fig. \ref{fig:Fig6_4}. However, Dimitriadis and Polemi (2006) \cite{53} proposed a protocol, named, IDM3G, to achieving the mutual authentication and identity privacy in 3G. The IDM3G protocol use two phases, namely, 1) the authentication of the UMTS Subscriber Identity Module (USIM) by providing a personal identification number and 2) the mutual authentication between the USIM and the mobile operator. By using the authentication request based on HTTP, the IDM3G is efficient in term of the number of messages exchanged in the path, which is lower compared to both protocols \cite{127} \cite{128}, but the location privacy is not considered. Similar to the IDM3G protocol, Dimitriadis and Shaikh (2007) \cite{54} proposed a protocol, called BIO3G, for establishing secure and privacy friendly biometric authentication in 3G mobile environments. The BIO3G protocol cannot resist against the DoS attacks and the location and identity privacy are not considered compared to the IDM3G protocol \cite{53}. He et al. \cite{59} proposed three categories of authentication scenarios for the 4G system. The main idea of \cite{59} is the use of \textit{Self-Certified Public-Key}, which need not be accompanied by a separate digital certificate. The advantage of the protocol \cite{59} is that it considers the identity privacy, but its disadvantage is the location privacy of mobile users. The following question is: Is it necessary to preserve the location privacy in future 5G networks? According to Lu et al. \cite{65}, ensuring location privacy in a cellular network is an effort to prevent any other party from learning the mobile users current and past locations. The recent idea in \cite{125} and \cite{299} can be applied for privacy preserving the social application under 4G/5G communications.

Location privacy is one of the most important models for privacy, as discussed in our previous surveys in \cite{155} \cite{156}. To the best of our knowledge, Lu et al. \cite{65} proposed the first study that deals with the mutual authentication with location privacy. Specifically, the authors proposed a novel mutual authentication protocol with provable link-layer location privacy. With the help of the \textit{Preset in Idle} technique, the protocol \cite{65} is efficient in terms of the packet delay time and the total packet time cost compared with the protocol \cite{167}. On the other hand, mutual authentication with identity privacy can also be preserved using the identity management mechanism proposed by Abdelkader et al. \cite{71}. The authors proposed an advanced Identity Management scheme, called AIM, in order to guarantee mutual authentication, privacy, and tracking avoidance for 4G networks. According to Saxena et al. \cite{116}, the EPS-AKA protocol of the LTE network does not support Internet of Things (IoT). Specifically, the authors proposed an authentication protocol for an IoT-Enabled LTE Network that is entirely based on the symmetric key cryptosystem.

For the security of future fifth generation telecommunications, a service provider will need to apply the managed security services (MSS) as network security services. According to Ulltveit-Moe et al. \cite{78}, the security services may be required for all mobile terminals such as antivirus, firewalls, Intrusion Detection Systems (IDS), integrity checking and security profiles. Specifically, the authors proposed a location-aware mobile intrusion prevention system with enhanced privacy, named mIPS, which is integrated into MSS. The mIPS system can preserve the personal privacy profile, but he needs to be evaluated in the future for 5G communications.  Using identification parameters, including, the International Mobile Subscriber Identity (IMSI) and the Radio Network Temporary Identities (RNTI), Jang et al. \cite{95} proposed an authentication protocol to safely transmit identification parameters in different cases of the initial attach under 4G mobile communications.

According to Madueno et al. \cite{276}, the LTE network is a promising solution for cost-efficient connectivity of the smart grid monitoring equipment. To ensure the security of this equipment, Haddad et al. \cite{102} proposed a privacy-preserving scheme to secure the communications of an automatic metering infrastructure via LTE-A networks. To share keys, the scheme \cite{102} uses a key agreement protocol between the smart meters, the utility company, and the LTE network. The scheme \cite{102} cannot only achieve the mutual authentication, key agreement, and key evolution but also can preserve the confidentiality/data integrity and authenticity. Recently, Mahmoud et al. \cite{114} proposed a privacy preserving power injection querying scheme over LTE cellular networks, to solve the problem of privacy exposure of storage unit owners. Therefore, the 4G/5G communications can be used by the traffic information systems \cite{103}. Gisdakis et al. \cite{103} addressed the security and privacy protection aspects of smartphone-based traffic information systems. More specifically, the authors proposed a privacy-preserving system using the architecture presented in \cite{275}. This system is based on three main phases, namely, 1) System initialization, 2) Device authentication and report submission, and 3) Device eviction. In addition, the system \cite{103} can provide the anonymity and the report unlinkability.
\begin{figure}
\centering
\includegraphics[width=0.6\linewidth]{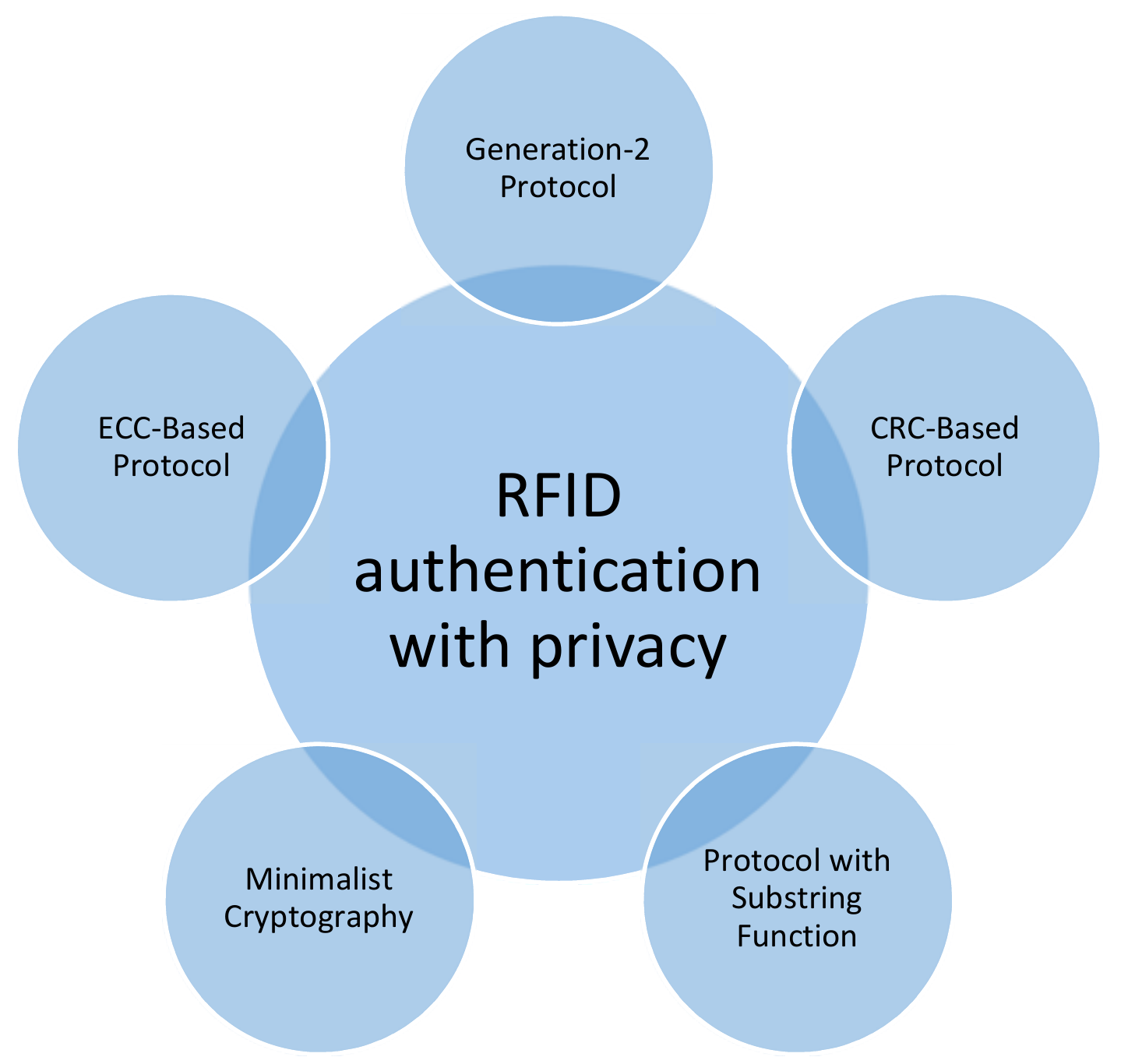}
\caption{Classification of RFID authentication protocols with RFID privacy}
\label{fig:Fig6_5}
\end{figure}
\subsection{RFID authentication with privacy}

Radio Frequency Identification (RFID) systems are low cost and convenience in identifying an object without physical contact, which consists of radio frequency (RF) tags, or transponders, and RF tag readers, or transceivers. According to Sun and Ting \cite{64}, RF technology can provide three functions: item awareness, information searching, and quality control. In addition, an RFID application contains three basic roles: 1) tag, 2) reader, and 3) back-end database \cite{163}. As presented in Fig. \ref{fig:Fig6_5}, RFID authentication protocols with RFID privac1) Generation-2 Protocol, e.g., the Gen2 protocol in \cite{138}, 2) CRC-Based Protocol, e.g., the SASI protocol in \cite{55}, 3) Minimalist Cryptography, e.g., the protocol in \cite{165}, 4) Protocol with Substring Function, e.g., the protocols in \cite{164} and \cite{166}, and 5) ECC-Based Protocol, e.g., the protocol in \cite{55}. Similair to Sun and Ting \cite{64}, Dubrova et al. \cite{108} proposed a message authentication scheme based on Cyclic Redundancy Check (CRC) codes for 5G Mobile Technology.

Chien \cite{55} proposed an ultralightweight RFID authentication protocol, called SASI, to providing strong authentication and integrity protection. The SASI protocol uses only simple bit-wise operations on the tag. Chien and Chen \cite{56} addressed the weaknesses of two schemes \cite{137} \cite{138} and proposed a mutual authentication scheme for GEN-2 RFID. The scheme \cite{56} can preserve the privacy and resist against DOS attack compared to both schemes \cite{137} \cite{138}. Liu and Bailey proposed an another interesting protocol that can achieve both privacy and authentication in \cite{63}. Specifically, the authors proposed a privacy and authentication protocol for passive RFID tags, called PAP. The PAP protocol is based on four main phases, namely, \textit{In-store, Checkout, Out-store, }and\textit{ Return}. PAP can resist against replay attack, but vulnerable to some attacks such as desynchronization attack and tracing attack. The following question is: Is it really necessary to detect the tracing attack? According to Sun and Ting \cite{64}, with tracing attack, an adversary have both a "malicious active reader" and several "malicious passive" loggers. The authors proposed a solution, called ${{\rm Gen2}}^+$, for RFID application with focusing on the protection of UltraHigh Frequency (UHF) passive tags from malicious readers. The ${{\rm Gen2}}^+$scheme can detect the tracing attack, also efficient in terms of the number of rounds required, and the period of key update compared to three schemes \cite{55}, \cite{138}, and \cite{164}. 

To achieve RFID authentication with anonymity/untraceability, and even availability, Chien and Laih \cite{66} proposed a RFID authentication protocol based on Error Correction Codes (ECC) \cite{174}. The protocol \cite{66} can achieve mutual authentication between the tags and the reader based on the successful verification of the PRNG function applied on the secret key. The protocol \cite{66} is efficient in term of computation complexity compared to the protocol LMAP \cite{134}. According to Kulseng et al. \cite{72}, the lightweight solution such as LMAP \cite{134} has been either broken or weakened. In fact, the authors in \cite{72} proposed a protocol in which only the authenticated readers and tags can successfully communicate with each other. Then, they designed protocols that achieve secure ownership transfer in three-party and two party low-cost RFID systems, but theses protocols need to be examined using real hardware. Especially for detection of man-in-the-middle attack, Li et al. \cite{94} proposed an authentication protocol, named LCMQ, which is proved secure in a general man-in-the-middle model. The LCMQ protocol can achieve RFID authentication and also efficient in terms of the tag's computation, storage, and communication costs compared with traditional cryptographic primitives such as RSA, DSA, and SHA.

Furthermore, Zhou et al. \cite{73} proposed a lightweight anti-desynchronization RFID authentication protocol, which is suitable for the low-cost RFID environment. Based on the idea of \textit{Index-pseudonym}, the protocol \cite{73} cannot only ensure the privacy of the tag, but also provide the forward security, location privacy, integrity, and tag anonymity. The strong advantage of the protocol \cite{73} is in desynchronization resistance compared to the protocol \cite{72}. By using a modified EAP-AKA protocol \cite{234} for authentication with the access network, Sharma and Leung \cite{84} proposed a robust one-pass IMS authentication mechanism in LTE-fem to cell heterogeneous networks. The mechanism is 50 percent improvement over the existing multi-pass authentication scheme published before 2012. Liao and Hsiao \cite{99} proposed a secure ECC-based RFID authentication scheme integrated with ID-verifier transfer protocol, which is efficient in terms of computational cost and communication overhead compared to the scheme of Tuyls et al. \cite{260}.

To preserve the authentication for IoT in 5G. Fan et al. \cite{110} proposed a lightweight RFID mutual authentication protocol with cache in the reader, named LRMAPC. Using an ultralightweight RFID mutual authentication protocol with cache in the reader, the LRMAPC protocol can achieve mutual authentication and provide forward security. Recently, Sun et al. \cite{119} formulated secure and privacy preserving object finding via mobile crowdsourcing. Then, they proposed a scheme, called SecureFind. Based on the initial object-finding request, the SecureFind scheme can obtain the information the service provider. Based on the vulnerability of two published protocols RRAP \cite{231} and RCIA \cite{302}, Luo et al. \cite{126} proposed recently a new ultra-lightweight mutual authentication protocol, which doesn't use any unbalanced operations like \textit{OR} and \textit{AND}.
\begin{figure}
\centering
\includegraphics[width=1\linewidth]{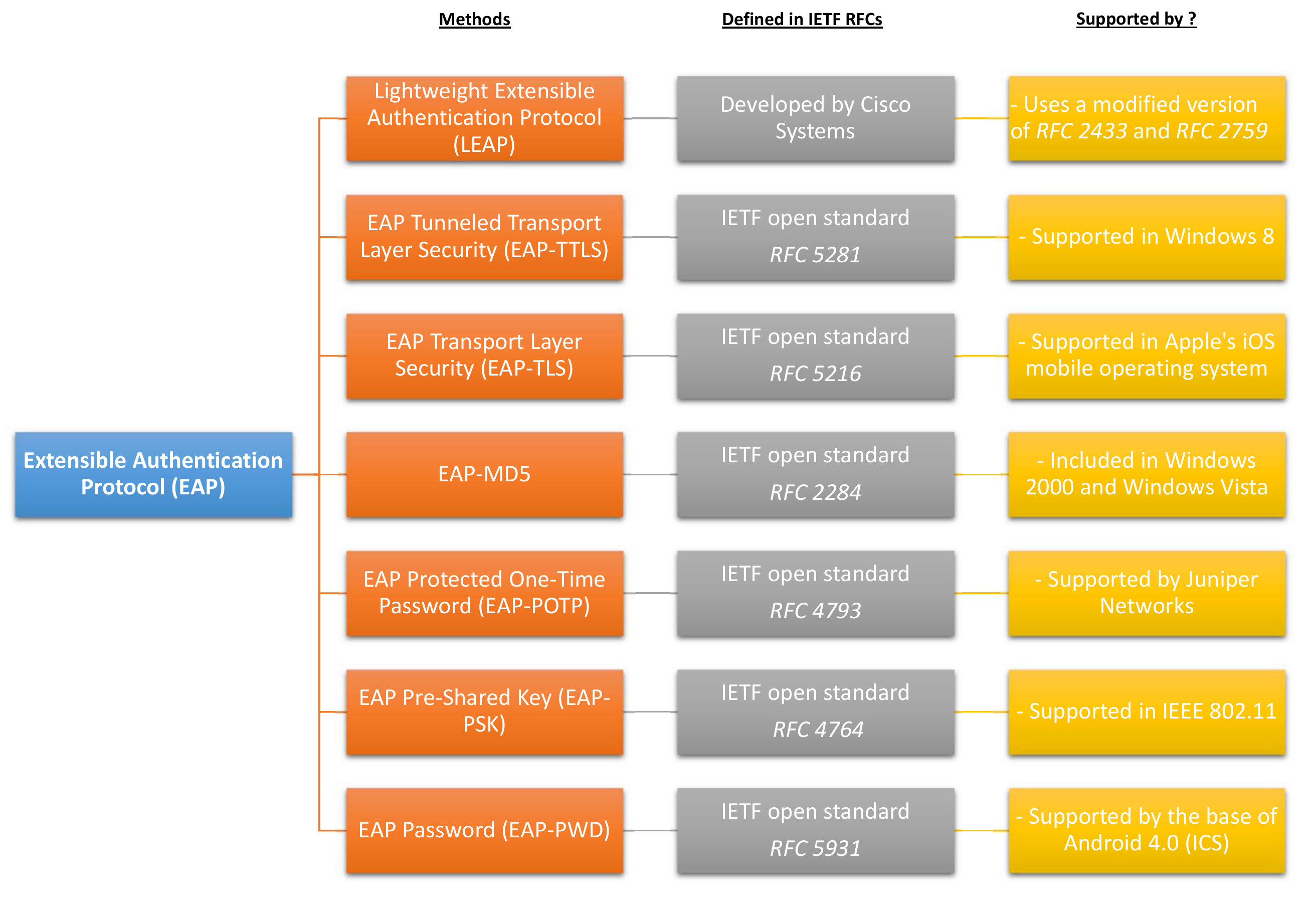}
\caption{Classification of methods based on EAP authentication framework}
\label{fig:Fig6_6}
\end{figure}
\subsection{Deniable authentication with privacy}

The deniable authentication differs from traditional authentication in a way that the Receiver cannot convince a third party \cite{141}. Therefore, Lee et al. \cite{57} proposed a protocol based on the non-interactive manner in order to achieve deniable authentication. Based on the shared session secret and the ElGamal signature scheme \cite{140}, the protocol \cite{57} does not only consider the security issues proposed in \cite{142}, including forgery attack, impersonation attack, deniability, and completeness but can also sustain the security when the session secret has already been compromised. Therefore, the use of message authentication codes (MACs) \cite{131} between two parties in cellular networks can achieving the deniable authentication.

To providing a lower degree of scalability and security, Bersani and Tschofenig \cite{58} defined an experimental protocol for the Internet community, called EAP-PSK, under the RFC 4764. The Extensible Authentication Protocol (EAP) is an authentication frequently used in wireless networks that defined in RFC 3748 \cite{148}, RFC 2284 \cite{149}, and was updated by RFC 5247 \cite{150}. As detailed in Fig. \ref{fig:Fig6_6}, there are many EAP authentication framework-based methods, which published as RFCs \cite{151} as Internet Standards. However, Chen et al. \cite{61} proposed two strong devices and user authentication schemes for Wi-Fi and WiMAX inter-networked wireless cities. The idea of \cite{61} is based on the modified Transport Layer Security (TLS) protocol \cite{158}, which leverage Trusted Platform Module (TPM) technologies \cite{159}. The work \cite{61} does not consider the identity and location privacy. Besides, the following question is: can we use the EAP to achieve the identity privacy? According to Pereniguez et al. \cite{70}, if the authentication mechanism does not have an adequate level of privacy, the identity and location can be revealed. Pereniguez et al. \cite{70} proposed a privacy-enhanced fast re-authentication, named 3PFH, for EAP-based 4G of mobile communications. The main idea of 3PFH is defined by a multi-layered pseudonym architecture to achieve user anonymity and untraceability. The 3PFH is applicable when the handoff takes place between different network operators. In addition, Arul et al. \cite{124} proposed a caching mechanism, called UPP-KC, where the keys are cached only along a predicted path for broadband wireless networks.

\subsection{Authentication with mutual anonymity}

Anonymity is an important security aspect of cellular communications, since it protects the privacy of the users, as discussed in our previous survey in \cite{155}. Lu et al. \cite{60} proposed an anonymous zero-knowledge authentication protocol, called PT, for Peer-to-peer (P2P) systems. We note here that we have selected this protocol because it can apply as an authentication protocol in 4G and 5G cellular communications. Besides, the PT protocol can support trust management in anonymous environments and scalable in both static and dynamic environments. To provide integrity to data exchanges after authentication, the PT protocol uses a Diffie-Hellman Key Exchange protocol into the authentication procedure to generate a session key. 

\begin{figure}
\centering
\includegraphics[width=0.7\linewidth]{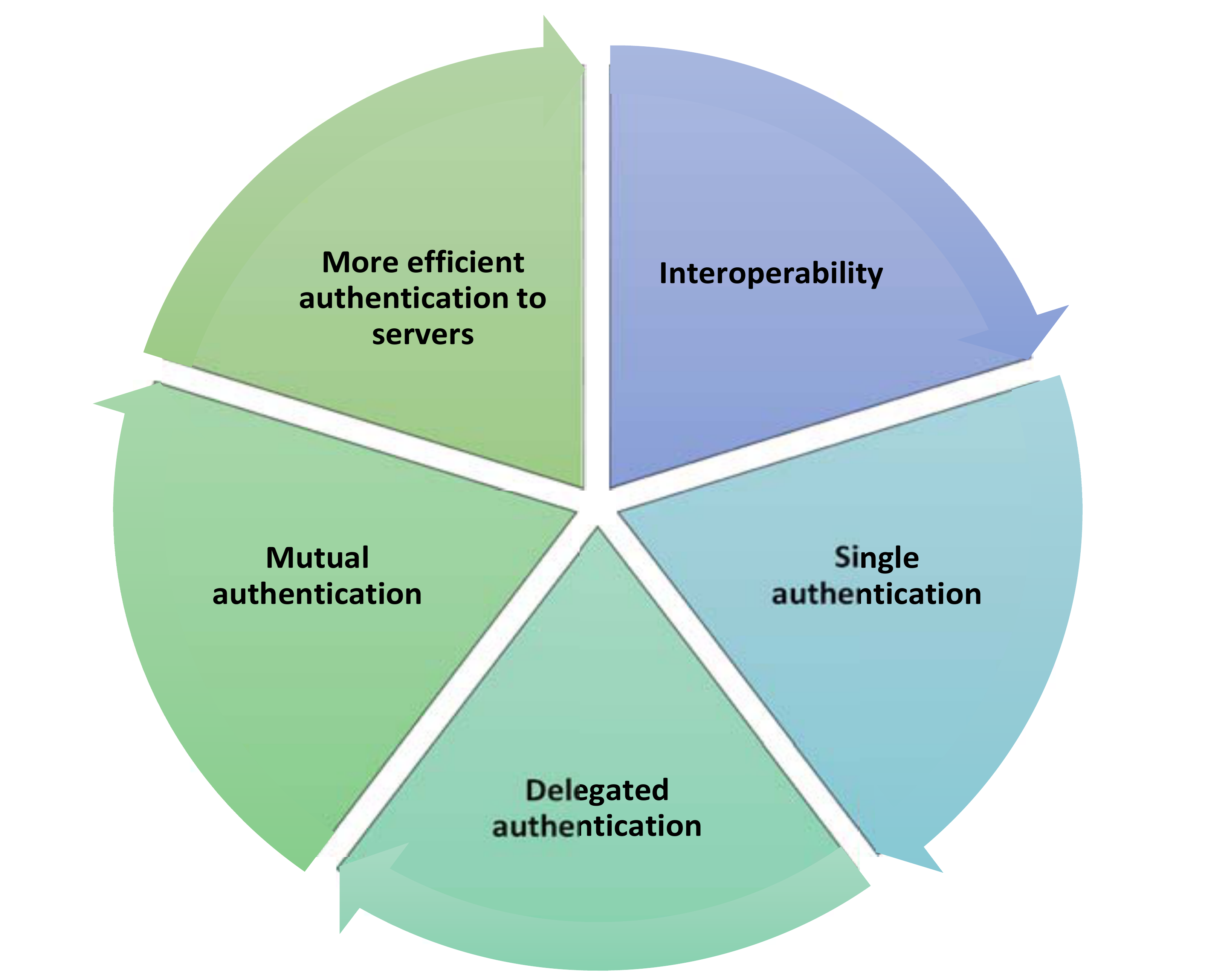}
\caption{Different models offered by the Kerberos protocol}
\label{fig:Fig6_7}
\end{figure}

To achieving the privacy preserving context transfer for 4G networks, Terzis et al. (2011) \cite{75} proposed four privacy preserving schemes for Context transfer protocol (CXTP) \cite{206}. These schemes are efficient in terms of application handoff service time compared to CXTP, while at the same time guarantee the privacy of the end-user. To verify the identity of a user or a host over 4G network, the network authentication protocol, called Kerberos, can be used. The Kerberos protocol is proposed under IETF RFC 4120 \cite{207}, where he composed with several entities, including, 1) The client $(C)$ with its own secret key, 2) The server $(S)$ with its secret key, 3) Ticket-granting service, and 4) Key distribution center.  As presented in Fig. \ref{fig:Fig6_7}, the Kerberos protocol provides several authentication models, including, 1) More efficient authentication to servers, 2) Interoperability, 3) Single authentication, 4) Delegated authentication, and 5) Mutual authentication. However, According to Pereniguez et al. \cite{77}, the Kerberos protocol suffers from two issues, namely, user anonymity and service access untraceability. The authors proposed a two-level privacy architecture, named PrivaKERB, to preserves the privacy of the user during activity with Kerberos. Based on two different levels of privacy: level 1, which provides user anonymity through pseudonyms, and level 2 where, apart from user anonymity, service access untraceability is assured. In addition, PrivaKERB is efficient in terms of service times, resource and network utilization compared to the standard Kerberos protocol \cite{207}.

Recently, Zhang et al. \cite{117} proposed a secure data sharing strategy for Device-to-Device (D2D) in 4G LTE-advanced network, called SeDS. To ensures data confidentiality, integrity, non-repudiation, and system availability, the SeDS strategy uses the digital signature and symmetric encryption. In addition, the SeDS strategy is efficient in terms of computational overhead, communication overhead and availability in a practical D2D communication environment. The idea of Hashem Eiza et al. \cite{118} can be applied for cloud-assisted video reporting service in 5G enabled vehicular networks.
\subsection{Authentication and key agreement with privacy}

The Authentication and Key Agreement (AKA) protocol is a challenge-response based mechanism that uses symmetric cryptography. The Universal Mobile Telecommunication System (UMTS) has adopted the AKA protocol of 3GPP, known as a standard of 3G with RFC 3310. Therefore, Deng et al. \cite{62} proposed an improved authentication and key agreement protocol based on public key cryptosystem. The protocol \cite{62} is vulnerable to some attacks, such as replay attack, man-in-the-middle attack, and DoS attack.  The following question is: Is it really necessary to hiding communication content from the external adversary under AKA protocol? Hamandi et al. \cite{120} proposed a hybrid scheme based on modifications to the LTE-AKA scheme, which employs both symmetric and asymmetric key encryption in order to detect and avoid both insider and outsider attacks. Using an efficient access-policy updating method, Li et al. \cite{121} proposed a group-based AKA protocol, called GR-AKA. The GR-AKA can reduce the communication overhead and alleviates the burden between machine type communication devices, but the known-key secrecy and the perfect forward secrecy are not considered compared to the scheme \cite{101}. To avoid the signal congestion in 3GPP networks, Yao et al. \cite{122} proposed a group-based authentication for machine-to-machine (M2M), which is efficient in term of bandwidth consumption.

According to Zhu et al. \cite{68}, the AKA protocol can easily be extended to provide revocable privacy by adopting the fair blind signature technique \cite{185}. Specifically, Zhu et al. \cite{68} proposed an anonymous authenticated key agreement protocol, called PPAB, to achieve scalable, authentication and billing in the context of interdomain roaming in the wireless metropolitan area sharing networks (WMSNs). The PPAB protocol considers five levels of privacy protection, namely, 1) \textit{content privacy}, 2) \textit{external privacy}, 3) \textit{internal privacy I}, 4) \textit{internal privacy II}, and 5) \textit{internal privacy III}. The \textit{content privacy} is hiding communication content from the external adversary. The \textit{external privacy} is hiding identity information of mobile users from the external adversary. The \textit{internal privacy I} is hiding identity information of mobile users from the wireless Internet service providers. The \textit{internal privacy II} is hiding identity information of mobile users from the roaming broker. The \textit{internal privacy III} is hiding identity information of mobile users from adversary for each handoff event \cite{68}. Besides, PPAB is efficient in term of energy consumption compared to the scheme \cite{184}, but the deniability and completeness are not considered. To the best of our knowledge, the work of Zhu et al. \cite{68} is the first study on the issues of localized authentication, billing, and privacy in the context of interdomain roaming in the WMSNs. To achieve the protection of user privacy, anonymity and untraceability for roaming network, Zhang et al. \cite{105} proposed a privacy-preserving authentication scheme based on elliptic curve cryptography.

The Session Initiation Protocol (SIP) is proposed by IETF under RFC 3261 in full and a number of extension RFCs including RFC 6665 (event notification) and RFC 3262 (reliable provisional responses). The SIP protocol is an IP-based telephony protocol for multimedia telecommunications (sound, image, etc.) in 3G mobile networks and over Internet Protocol (IP) networks. According to Wu et al. \cite{69}, the SIP protocol does not include any specific security mechanisms. Specifically, Wu et al. \cite{69} proposed a provably secure authentication and key agreement protocol for SIP using elliptic curve cryptography, called NAKE, in order to achieving the perfect forward secrecy. The NAKE protocol is preferable in the applications that require low memory and rapid transactions. However, the disadvantage of the NAKE protocol is that it does not preserve the location privacy compared to the scheme \cite{68}. To provide the location and identity privacy, Karopoulos et al. (2011) \cite{76} proposed two solutions in SIP, where the first the ID of the caller is protected while in the second both IDs of the caller and the callee are protected. Both solutions consider the identity privacy and are efficient in term of mean server response delays compared to standard SIP, but the key agreement is not considered.

To improve the security of both schemes, including, Wu et al.'s scheme \cite{236} and Yoon et al.'s scheme \cite{237}, He \cite{87} proposed a new user authentication and key agreement protocol using bilinear pairings for mobile client--server environment. The idea of \cite{87} is based on the bilinear pairing under the computational Diffie--Hellman (CDH) and collision attack assumption and in the random oracle model. The scheme \cite{87} can achieve the client-to-server authentication, the server-to-client authentication and key agreement under the random oracle model, but the privacy is not considered compared to the scheme \cite{82}. However, using a temporary confidential channel, Chen et al. \cite{91} designed three type of authentication, including, 1) Bipartite authentication protocol, 2) Tripartite authentication protocol, and 3) Multipartite transitive authentication.

Based on three main categories of auxiliary channels, including, \textit{input}, \textit{transfer}, and \textit{verification}, Mayrhofer et al. \cite{93} proposed a unified auxiliary channel authentication protocol, named UACAP, which releases a specific implementation in the form of the Open-source Ubiquitous Authentication Toolkit (OpenUAT) \cite{248}. Using two main phases, namely, 1) Diffie-Hellman key exchange with precommitment and 2) Out-of-band key verification, the UACAP protocol can exploit any combination of security guarantees from arbitrary auxiliary channels. Recently, Ramadan et al. \cite{115} proposed a user-to-user mutual authentication and key agreement scheme, which is more compatible with the LTE security architecture. The scheme \cite{115} is based on four phases, namely, 1) Setup and key generation, 2) Authentication between the users and the mobility management entity, 3) User-to-User authentication, and 4) Establish a shared secret key.
\begin{figure}
\centering
\includegraphics[width=0.7\linewidth]{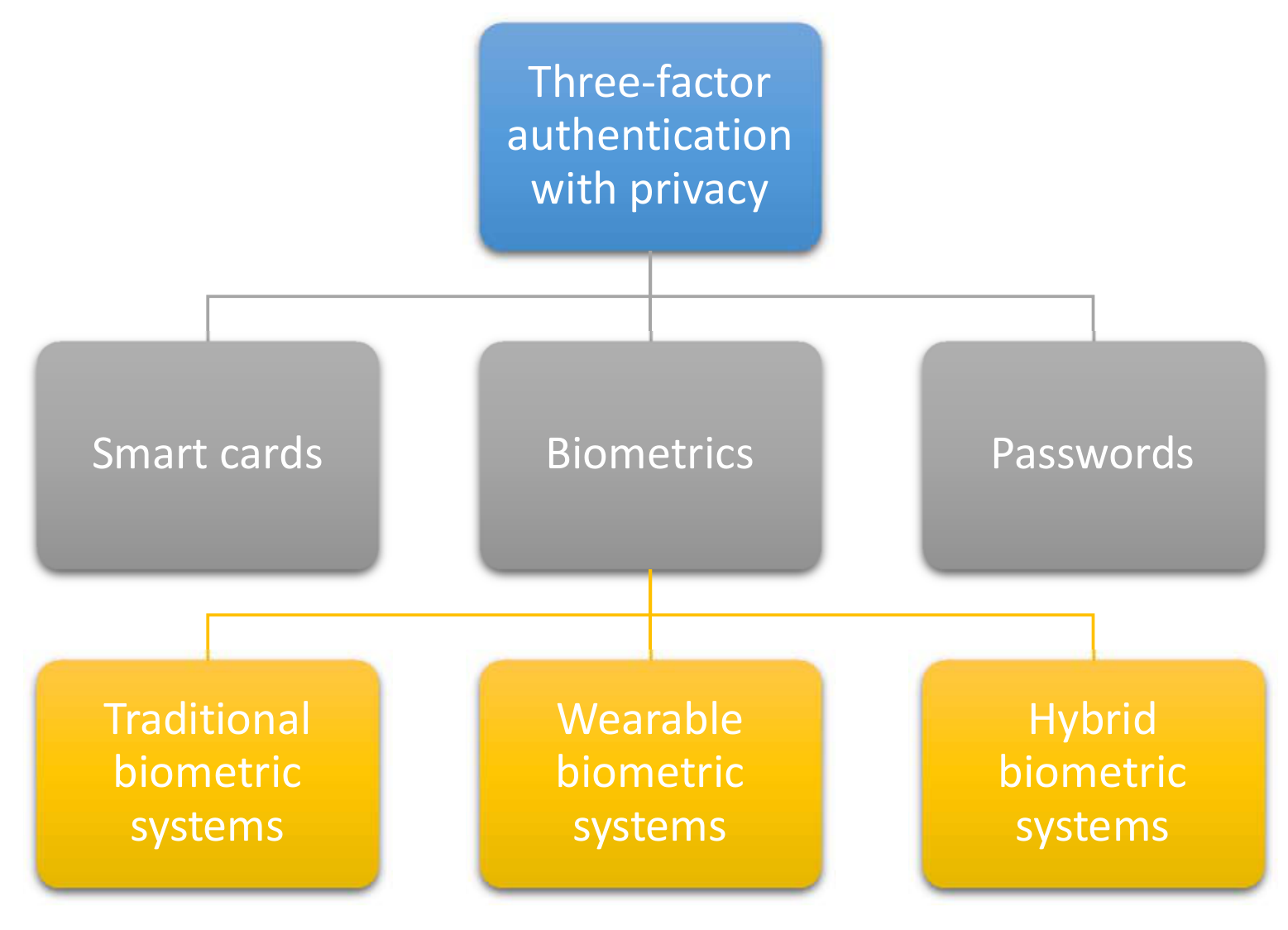}
\caption{Classification of three-factor authentication schemes with privacy}
\label{fig:Fig6_8}
\end{figure}
\subsection{Three-factor authentication with privacy}

The three-factor authentication schemes with privacy can mainly be classified into three categories: 1) Smart cards-based protocol, 2) Passwords-based protocol, and 3) Biometrics-based protocol, as presented in Fig. \ref{fig:Fig6_8}. The following question is: can we use the three factors together? According to Fan and Lin \cite{67}, this three different data types can be used together in an authentication protocol, where smart cards show \textit{what you have}, passwords represent \textit{what you know}, and biometrics mean \textit{what you are}. Specifically, the authors proposed a truly three-factor authentication scheme to achieving the strong biometrics privacy.  Based on the login and authentication phase, the server accepts only if each factor (password, smart card, and biometric data) passes the authentication. The protocol \cite{67} is efficient in term of low computation for smart cards compared to three-factor authentication schemes in \cite{175} and \cite{176}. Therefore, according to Blasco et al. \cite{196}, the biometric systems can mainly be classified into three categories: 1) Traditional Biometric Systems (e.g., Windows Hello \cite{198}), 2) Wearable biometric systems (e.g., Using a smartphone), and 3) Hybrid biometric systems (e.g., Hybrid systems arises in telecare services \cite{200}). For more details in the field of wearable biometrics and in the security and privacy issues in implantable medical devices, we refer the reader to the both recent surveys \cite{196} and \cite{197}. 

For security of 4G and 5G networks using Biometric-based identification, it is required that the client does not learn anything on the database. Therefore, according to Barni et al. \cite{74}, the fingerprint is likely to be used in applications that need higher reliability. Specifically, the authors proposed a privacy-preserving system for fingerprint-based authentication. Based on the Fingercode representation introduced in \cite{205}, the identification protocol \cite{74} is efficient in term of bandwidth usage compared to both schemes Erkin et al. \cite{201} and Sadeghi et al. \cite{202}. Especially for multiserver environment, He and Wang \cite{111} proposed a biometrics-based authentication scheme, which is overcome the weaknesses in Yoon and Yoo's scheme \cite{288} and Kim et al.'s scheme \cite{289}.

The password-based-authentication protocols are a reliable solution to provide identity protection and satisfy strong mutual authentication in 4G and 5G networks. Moreover, Sood et al. \cite{79} presented a cryptanalysis of the Hsiang and Shih protocol \cite{211} in order to propose a secure dynamic identity-based authentication protocol for multi-server architecture. The protocol \cite{79} is efficient in term of computation complexity compared to related smart card based multi-server authentication protocols \cite{211} \cite{212}. Similar to \cite{79}, Lee et al. \cite{80} presented a cryptanalysis of Hsiang et al. scheme \cite{211} where the authors have found that Hsiang et al. scheme is still vulnerable to a masquerade attack, server-spoofing attack, and is not easily reparable. Then, the authors \cite{80} proposed a scheme to solve these weaknesses. In addition, Lee et al. scheme \cite{80} is efficient in term of communication cost of the login and verification phase compared to three schemes, namely, Hsiang et al. scheme \cite{211}, Liao-Wang scheme \cite{212}, and Chang-Lee scheme \cite{213}. Li et al. \cite{85} out that the protocol \cite{79} is still not secure. Based on the cryptanalysis of the protocol \cite{79}, the authors \cite{80} proposed a security dynamic identity-based authentication protocol for multi-server architecture. The protocol \cite{85} provide the user's anonymity, proper mutual authentication, and session key agreement. In addition, the protocol \cite{85} is efficient in terms of computational complexity compared with some related dynamic ID based multi-server authentication protocols, including, \cite{79}, \cite{211}, and \cite{212}. Similar to \cite{91}, Liu and Liang \cite{92} proposed a hierarchical identity-based access authentication protocol, named HA-HIBS-VN, which can be applied for 4G and 5G cellular networks. The HA-HIBS-VN protocol \cite{92} can provide the private key privacy and signature unforgeability. By combining the peer group tree (PGT), identity-based signature, and designed mobile vector network protocol (MVNP), the HA-HIBS-VN protocol \cite{92} is efficient in term of handover delays compared with the protocol in \cite{244}.

Previous works in this area, i.e., password-based-authentication, have come short of providing solutions to detecting password reuse attacks. Moreover, Sun et al. \cite{86} proposed the first user authentication protocol to prevent password stealing (i.e., phishing, keylogger, and malware) and password reuse attacks simultaneously. The authors proposed a user authentication, called oPass, to thwart this both attacks. The idea of oPass is to adopt one-time passwords, which they expired when the user completes the current session. Based on two main processes, including, 1) Login phase and 2) Recovery phase, the oPass can protect the information on the cellphone from a thief. Hence, the oPass protocol can be applied for 4G and 5G cellular networks. To provide privacy-preserving and secure roaming service for 4G and 5G cellular networks, Wang et al. \cite{96} revisited the privacy-preserving two-factor authentication scheme presented by Li et al. in \cite{252}, which they showed that the scheme \cite{152} suffers from offline password guessing attack. The study of Wang et al. \cite{96} can withstand offline password guessing attack even if the victim's smart card is lost. As an additional benefit, the Wang et al. scheme \cite{96} is efficient in term of computation cost on user side compared to fives schemes Li et al. \cite{252}, Isawa-Morii \cite{253}, He et al. \cite{254}, Zhou-Xu \cite{255}, and Xu et al. \cite{256}. To overcome the weaknesses of Das scheme \cite{269}, Li et al. \cite{101} a three-factor remote user authentication and key agreement scheme using biometrics. Based on discrete logarithm problem \cite{271}, the Li et al. scheme \cite{101} can ensure the known-key secrecy and provide the perfect forward secrecy.

Similar to \cite{96}, Chen et al. (2014) \cite{97} proposed an improved smart-card-based password authentication and key agreement scheme. Based on the review of the Xu et al. scheme \cite{258} and Sood et al. scheme \cite{257}, the scheme \cite{97} can achieve mutual authentication and guarantees forward secrecy. In addition, the scheme \cite{97} is efficient in term of computation cost on server side compared to three schemes Xu et al. \cite{258} Sood et al. \cite{257}, and Song \cite{182}. Therefore, if the authentication stored in the memory device is exposed, Jiang et al. pointed out that the scheme \cite{97} suffers from offline password guessing attacks.

Based on the cryptanalysis of two schemes of Chen et al. \cite{79} and Wen--Li \cite{259}, Ma et al. \cite{98} proposed three principles designing more robust schemes in the future, which can be applied for 4G and 5G cellular networks. To overcome the weaknesses of Chang et al.'s scheme \cite{268}, Kumari et al. \cite{100} proposed an improved user authentication scheme with session key agreement facility. Compared to the Chang et al.'s scheme \cite{268}, the Kumari et al.' scheme \cite{100} cannot only provide forward secrecy, but the user is anonymous and untraceable. To overcome the weaknesses of Kumari et al.'s scheme \cite{277}, Chaudhry et al. \cite{104} proposed an enhanced remote user authentication scheme, which can ensures privacy and anonymity.

\begin{figure}
\centering
\includegraphics[width=1\linewidth]{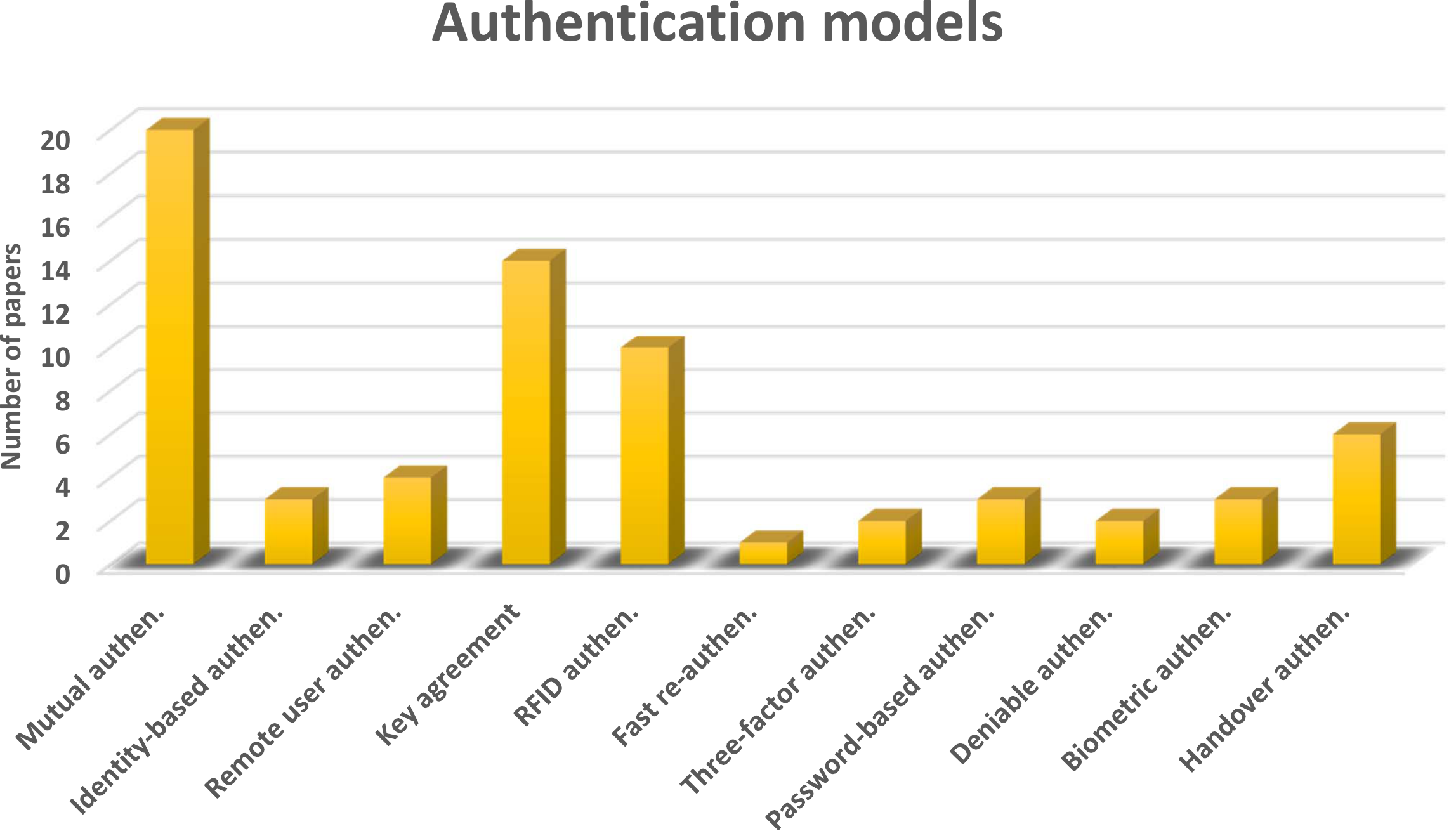}
\caption{Number of papers vs. Authentication models}
\label{fig:Fig7_a}
\end{figure}
\begin{figure}
\centering
\includegraphics[width=0.8\linewidth]{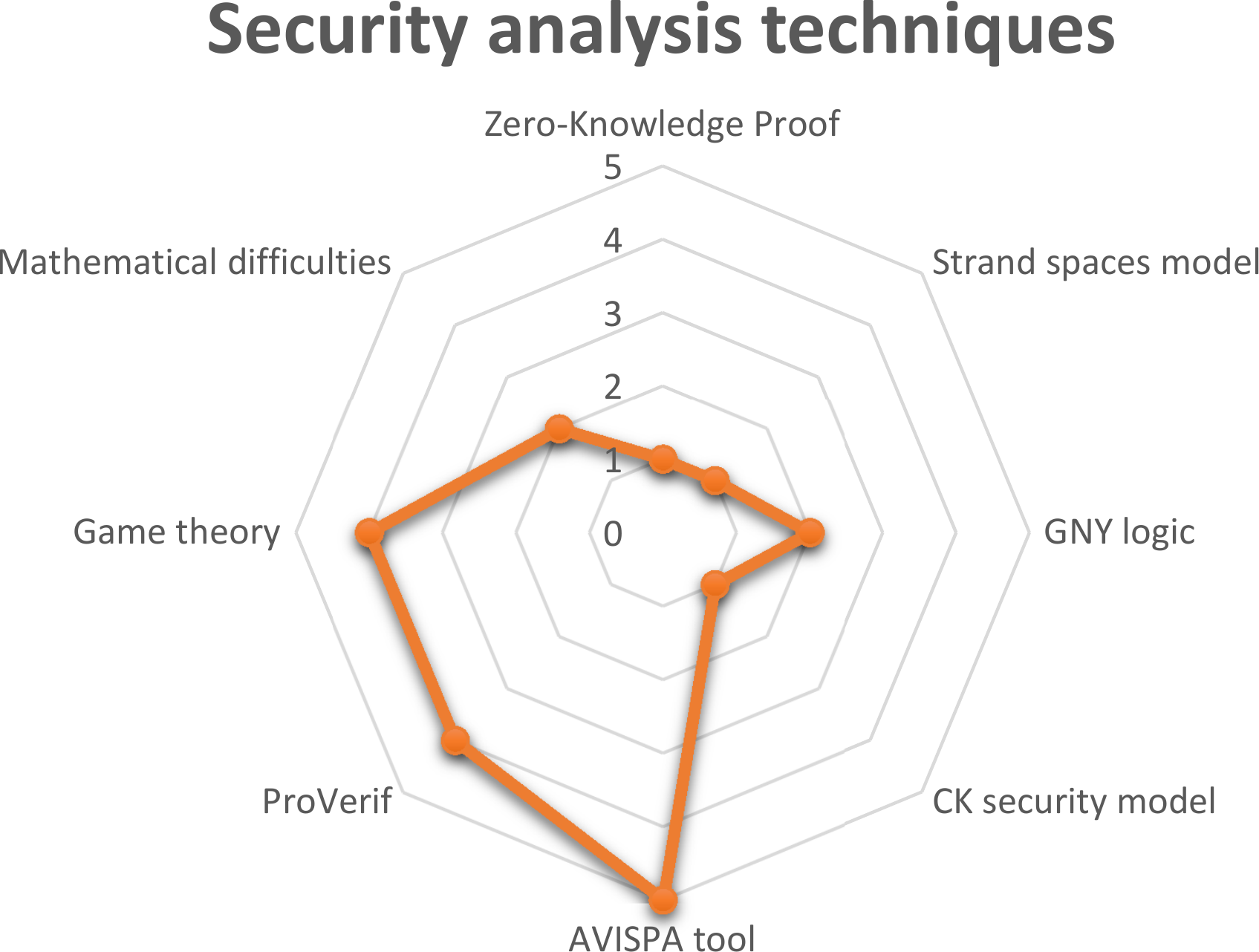}
\caption{Number of papers vs. Security analysis techniques}
\label{fig:Fig7_c}
\end{figure}
\begin{figure}
\centering
\includegraphics[width=1\linewidth]{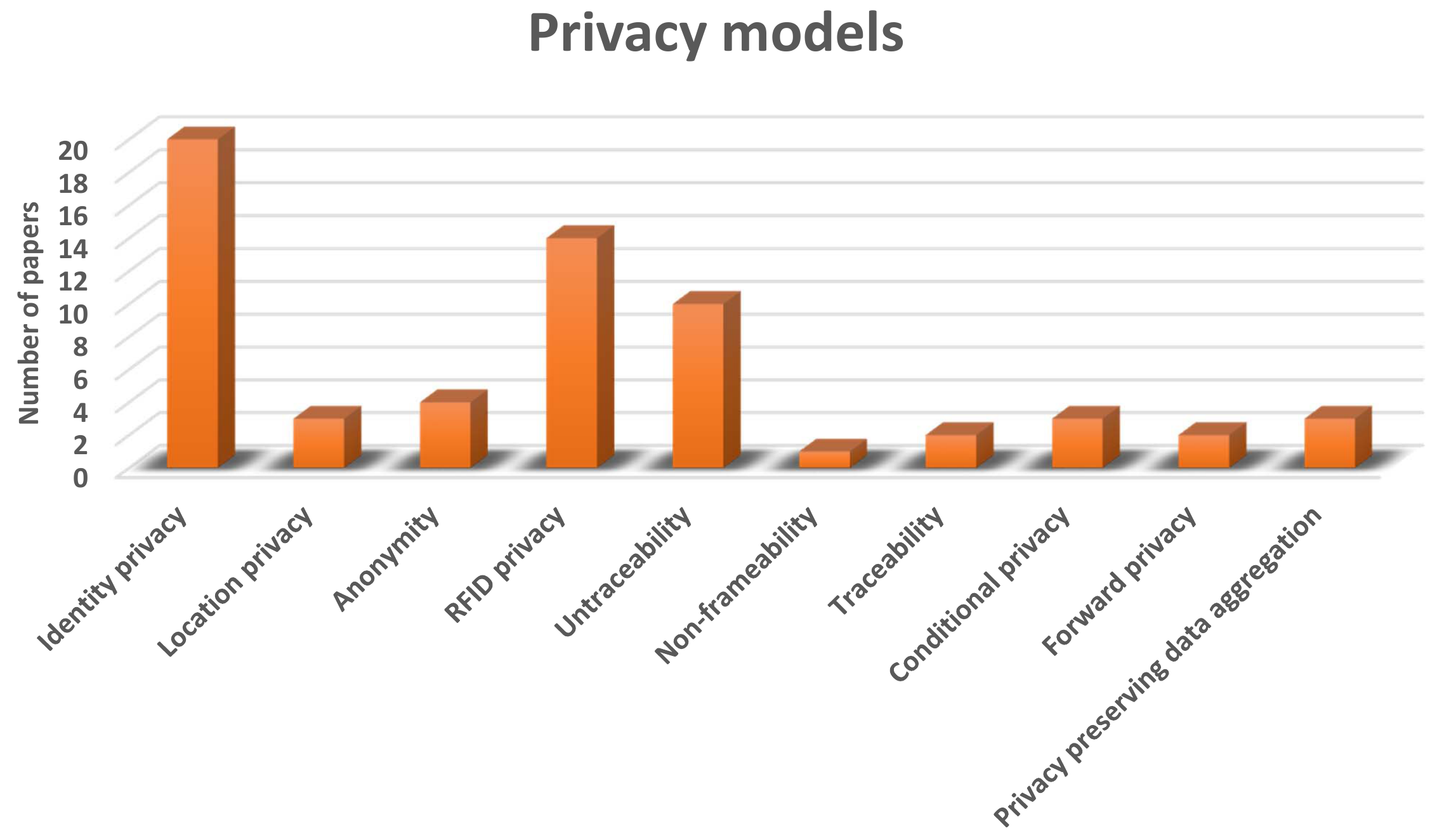}
\caption{Number of papers vs. Privacy models}
\label{fig:Fig7_b}
\end{figure}

\section{Future directions}\label{sec:future-directions}

As shown in Fig. \ref{fig:Fig7_a}, authentication and privacy-preserving schemes for 4G and 5G cellular networks focus on four authentication models, namely, proper mutual authentication, session key agreement, RFID authentication, and handover authentication. When security analysis techniques are used, the surveyed schemes use AVISPA tool, ProVerif, Game theory, and GNY logic, as shown in Fig.\ref{fig:Fig7_c} . In addition, the surveyed schemes focus on different areas of privacy models, namely, identity privacy, RFID privacy, untraceability, anonymity, and location privacy, as shown in Fig.\ref{fig:Fig7_b}. To complete our overview of authentication and privacy-preserving schemes for 4G and 5G cellular networks, open directions for future research are described below.

\subsection{Privacy preservation for Fog paradigm-based 5G radio access network}
To meet the requirements of mission-critical applications in 5G radio access network (RAN), two system design paradigms can be used, including, cloud and fog. Recently, Ku et al. \cite{L8} proposed a Fog-cloud integrated RAN architecture, named F-RAN. To integrate the computing functionality to 5G cellular networks, F-RAN adopts two approaches, including, loosely-coupled and tightly-coupled. However, several attacks against privacy can be launched, such as man-in-the-middle attack and replay attack, which can reveal the location and identity of Fog nodes in F-RAN architecture. How to provide the location and identity privacy for Fog paradigm-based 5G radio access network? Hence, privacy preservation for Fog paradigm-based 5G radio access network should be exploited in the future. 
\subsection{Authentication for 5G small cell-based smart grids}
The smart grid deployment requires a faster communication medium in the long run, which can be achieved by the 5G wireless from data and control plane isolation. With evolved multimedia broadcast and multicast communication, between aggregators (small cells) and smart grid consumers, Saxena et al. \cite{L9} introduced a planning of 5G small cells, for optimal demand response in smart grids. This idea can reduce energy production cost by 30\%, but this is calculated without taking in mind possible network attacks that also affect energy consumption. Even though numerous authentication schemes have been designed in recent years to protect communication in smart grids but these schemes are not reliable to detect and prevent common attacks as well as reducing energy production cost for 5G small cell-based smart grids. Therefore, how do we reduce the cost of energy under network attacks? Hence, authentication for 5G small cell-based smart grids is another possible future direction.
\subsection{Privacy preservation for SDN/NFV-based architecture in 5G scenarios}
Software Defined Networking (SDN) and Network Function Virtualization (NFV) technologies considered as key drivers to paving the way towards the 5G era, as discussed in a recent survey published in 2017 \cite{L10}. Specifically, Nguyen et al. reviews the state of the art of SDN/NFV-based mobile packet core network architectures, none of them carries study for the privacy preservation. A possible research direction in this topic could be related to privacy preservation for SDN/NFV-based architecture in 5G scenarios such as location privacy, identity privacy, anonymity, etc. Last but not least, guaranteeing the authentication between the mobile users and devices are also important factors when realizing the network sharing based on SDN/NFV in 5G scenarios.
\subsection{Dataset for intrusion detection in 5G scenarios}
As we have seen in subsection \ref{sec:countermeasures}, data mining and machine learning methods are used to help discover, determine, and identify unauthorized use and destruction of information systems, such as 4G and 5G cellular networks \cite{L11}. Buczak and Guven \cite{L11} have found that the most intrusion detection systems used the DARPA 1998, DARPA 1999, DARPA 2000, or KDD 1999 data sets. Therefore, the question we ask here is: Can these data sets be used in 5G scenarios? In other words, the threat models discussed in subsection \ref{sec:threat-models} are simulated in these data sets? We believe that further research is needed to develop a new data set to build a network intrusion detector under 5G environment.

\subsection{Privacy preserving schemes for UAV systems in 5G heterogeneous communication environment}
In a connected society, i.e., IoT, the intelligent deployment of Unmanned Aerial Vehicle (UAVs) in 5G heterogeneous communication environment will play a major role in improving peoples' lives. With the limitation of wireless communication and computing capability of drones, the application of UAV is becoming more and more complicated, especially for security and privacy issues. In a work published in 2017, He et al. \cite{L13} categorized threat models on the drone-assisted public safety network, in four categories, namely, attacks on confidentiality, attacks on integrity, attacks on availability, and attacks on authenticity. One possible future direction is to develop a privacy preserving schemes for UAV systems in 5G heterogeneous communication environment.

\subsection{Authentication and privacy-preserving schemes for 5G small cell-based vehicular crowdsensing}
Vehicular crowdsensing entails serious security and privacy issues, where it is important to protect user identity, location privacy, among others. Using Fog computing, Basudan et al. \cite{L20} proposed a new idea for privacy preserving for vehicular crowdsensing. Specifically, this idea introduced a certificateless aggregate signcryption scheme, named CLASC, which is highly efficient in term of low communication overhead and fast verification. Moreover, the system model considers that the road surface condition monitoring system comprises a control center, vehicles, smart devices, roadside units, and cloud servers. Since we are moving to the 5G communications, a new emerging paradigm will appear, named 5G small cell-based vehicular crowdsensing, in order to meet the requirements for new applications in vehicular ad hoc networks such as parking navigation, road surface monitoring, and traffic collision reconstruction. The future works addressing the limitations of authentication and privacy-preserving schemes for vehicular crowdsensing will have an important contribution for 5G small cell-based vehicular crowdsensing.

\section{Conclusions and lessons learned}\label{sec:conclusions}

In this article, we surveyed the state-of-the-art of authentication and privacy-preserving schemes for 4G and 5G cellular networks. Through an extensive research and analysis that was conducted, we were able to classify the threat models in cellular networks into attacks against privacy, attacks against integrity, attacks against availability, and attacks against authentication. In addition, we were able to classify the countermeasures into cryptography methods, humans factors, and intrusion detection methods. For the cryptographic methods, the surveyed schemes use three types of cryptographic, including, public-key cryptography, symmetric-key cryptography, and unkeyed cryptography. To ensure authentication, the surveyed schemes use three factors, including, what you know (e.g., passwords), what you have (e.g., smart cards), and 3) who are you (e.g., biometrics). For intrusion detection methods, the surveyed schemes use three systems, including, signature-based system, anomaly-based system, and hybrid IDS.

From security analysis point, there are twelve informal and formal security analysis techniques used in authentication and privacy preserving schemes for 4G and 5G cellular networks, namely, zero-knowledge proof, mathematical difficulties, GNY logic, CK security model, random oracle model, game theory, probabilistic functions, BAN logic, AVISPA tool, Open-source MIT Kerberos, OpenUAT, and ProVerif. We were able to classify these techniques in two classes, including, without an implementation tool and with an implementation tool.

Based on the categorization of authentication and privacy models, we were able to classify the surveyed schemes in seven types, including, handover authentication with privacy, mutual authentication with privacy, RFID authentication with privacy, deniable authentication with privacy, authentication with mutual anonymity, authentication and key agreement with privacy, and three-factor authentication with privacy.

Based on the vision for the next generation of connectivity, we proposed six open directions for future research about authentication and privacy-preserving schemes, namely, Fog paradigm-based 5G radio access network, 5G small cell-based smart grids, SDN/NFV-based architecture in 5G scenarios, dataset for intrusion detection in 5G scenarios, UAV systems in 5G environment, and 5G small cell-based vehicular crowdsensing.
%
\bibliographystyle{IEEEtran}
\bibliography{bare_jrnl_comsocd}

\end{document}